\documentclass[amsmath,
 amssymb,
 aps,
 nofootinbib,
 preprintnumbers
]{revtex4-1}
\usepackage[utf8]{inputenc}
\usepackage{simplewick}
\usepackage{footnote}
\usepackage{graphicx}
\usepackage{dcolumn}
\usepackage{bm}
\usepackage{stmaryrd}
\usepackage{amsmath}
\usepackage{physics}
\usepackage{amssymb}
\usepackage{mathtools}
\usepackage{bbm}
\usepackage{amsfonts}
\usepackage{xcolor}
\usepackage{footmisc}
\usepackage{comment}
\usepackage[normalem]{ulem}
\usepackage{slashed}
\usepackage{multirow}
\usepackage{appendix}
\usepackage{float}
\usepackage{braket}
\usepackage{hyperref}
\usepackage{xcolor}
\definecolor{cite}{rgb}{0.,0.,0.9}
\hypersetup{colorlinks,linkcolor={cite},citecolor={cite},urlcolor={cite}}
\usepackage[capitalise]{cleveref}
\usepackage{cancel}

\allowdisplaybreaks

\begin{document}
\preprint{ADP-26-14/T1311}
\title{Radion/Dilaton portal freeze-in dark matter in extra-dimensional theories}

\author{Amrita Mukherjee$^{a}$}
\author{George Sanamyan$^{b}$}
\author{Dipan Sengupta$^{a}$}

\affiliation{$^a$Sydney Consortium for Particle Physics and Cosmology, School of Physics,\\ The University of New South Wales, Sydney NSW 2052, Australia}
\affiliation{$^{b}$ARC Centre of Excellence for Dark Matter Particle Physics, Department of Physics, School of Physics, Chemistry and Earth Sciences, Adelaide University, South Australia 5005, Australia}

\begin{abstract}
We perform a comprehensive analysis of a radion/dilaton portal freeze-in dark matter in models of stabilized warped extra-dimensions. We show that the relic abundance of a generic dark matter particle (scalars, vectors, or fermions) is obtained via the conformal coupling of the radial mode in a consistent model stabilized warped extra-
dimensions; our calculations show that in scenarios where the reheating temperature is much smaller than the radion mass, keV-TeV scale dark matter particles can saturate the relic density.  Furthermore, we show that the model is subject to Lyman-$\alpha$ constraints in the keV–GeV range, providing a unique, model-independent probe of this sector. Our calculations extend previous attempts to calculate the freeze-in relic abundance in extra-dimensional portals by providing a consistent theoretical calculation of matrix elements, clearly laying out the underlying motivation and assumptions and providing a comprehensive account of the cosmological constraints on the model.

\end{abstract}

\maketitle

\tableofcontents

\vfill\eject

\section{Introduction}
\label{sec:intro}


Cosmological and astronomical observations indicate that the energy budget of the Universe contains a mysterious non-baryonic matter component comprising about $27\%$ of the total energy density. The influence of this {\it{invisible}} component is gleaned via its gravitational effects on the visible energy density in galaxies and galaxy clusters, as well as on the cosmic microwave background (CMB)~\cite{Planck2018,Bertone2016,deSwart2017}. If such a matter component originates from a fundamental physics source with a unique particle associated with it, then one has to look for physics beyond the standard model, since there are no standard model particles that can successfully act as a dark matter candidate.

The canonical mechanism by which a dark matter particle within a BSM theory saturates the relic density is via the freeze-out mechanism, where the dark matter particle, in equilibrium with the background SM thermal bath, annihilates via weak interactions to SM particles until it falls out of equilibrium due to the Hubble expansion of the Universe, leading to a frozen-out density of dark matter that we observe today. This ``so-called'' WIMP miracle predicts particles of masses $10$ GeV-few TeV, the sweet spot where a new BSM theory is also expected to solve the electroweak hierarchy problem. The lack of observation of WIMPs at direct and indirect detection probes, as well as at collider experiments such as the LHC, has placed severe constraints on leading dark matter theories, such as weak-scale supersymmetry and simple extensions of the standard model via additional fields. Therefore, alternative dark matter candidates such as axions or primordial black holes, as well as alternative mechanisms such as {\it{freeze-in}} \cite{Hall:2009bx} and particle asymmetry, have emerged as attractive propositions ( See \cite{Cirelli:2024ssz,Bertone2016} for comprehensive reviews).

We therefore remind ourselves that the dark matter problem arose from Zwicky's observation of a mismatch between observed galactic rotation curves and those predicted by Newtonian gravity~\cite{Zwicky:1937zza}. As all the primary evidence for dark matter stems from its gravitational interactions, it is reasonable to consider that a fundamental particle dark matter candidate interacts solely with the standard model (or beyond the standard model) through the gravitational force, leaving it inaccessible to particle physics experiments. There are a few possibilities here, 

\begin{enumerate}
    \item The inflaton, responsible for cosmic inflation post Big Bang, can produce dark matter via gravitational interactions~\cite{Barman:2021ugy} \footnote{The inflaton can also directly decay to dark matter via feeble interactions~\cite{Barman:2022tzk}, but this requires non-gravitational feeble interactions that couple the inflaton to dark matter, and goes beyond minimality.} at the end of the inflationary dynamics. This scenario can be realised either in standard cosmology or in modified cosmologies with early matter domination~\cite{Bernal:2019mhf}. Regardless, this depends on an underlying inflationary scenario and the inflaton dynamics. 
    \item Dark Matter can interact with the standard model via the massless spin-2 graviton that carries Einstein's theory of general relativity. This has been considered in the literature~\cite{Bernal:2018qlk}, and we will revisit this possibility. In addition, there have been studies considering simplified spin-2 portal models with freeze-in production of dark matter~\cite{Bernal:2018qlk}. There are unitarity constraints in this scenario that render the calculation questionable, especially when considered within a larger UV-complete model~\cite{Chivukula:2020hvi,Gill:2023kyz,Chivukula:2019rij,Chivukula:2023qrt,Chivukula:2022tla,Chivukula:2024nzt}.
    \item Dark Matter can interact gravitationally with the standard model via theories of compact extra-dimensions (warped or flat background), which predict an infinite tower of massive spin-2 particles (in addition to the massless graviton) and a scalar dilaton/radion particle.
\end{enumerate}

In all of the above, the dark matter particle is taken to be a standard model singlet and does not rely on any additional interactions.
While the possibility of inflaton interactions is quite attractive, one must take into account the inflaton's dynamics and nature. In this paper, we focus on the latter two possibilities, in which dark matter interacts with the standard model solely through gravity. 
The second scenario leads to a rather simple model in which interactions between dark matter and the standard model are mediated by the graviton. Over the years, numerous studies of such models using the number density formalism have shown that, due to the relative weakness of the gravitational interaction, achieving the correct relic density via an ultraviolet {\it{freeze-in}} mechanism in this scenario requires a very high reheating temperature~\cite{Garny2016,Tang2016,Garny2018,Tang2017,Bernal2018}. We will briefly go through this scenario.

One can extend the {\it{gravitational freeze-in}} scenario to include a mass for a hypothetical spin-2 mediator particle. Such mass terms (Fierz-Pauli mass terms) arise either in effective theories of gravity, such as de-Rham-Gabadadze-Tolley (dRGT)/bi-gravity~\cite{deRham:2014zqa}, or as a truncation to a single mode of a Kaluza-Klein decomposition of a higher-dimensional field. For the former, extra polarisations from the massive spin-2 state can naively enhance the high-energy scattering amplitudes of spin-2 particles to dark matter as $\frac{s^{3}}{m_{g}^{4}M_{\text{Pl}}^{2}}$, where $m_{g}$ is the mass of the massive spin-2 particle, and  $s$ is the centre of momentum energy, and $M_\text{PL}$ is the Planck mass. This would then naively imply that for low reheating temperature relative to the mass of the massive spin-2 particle, the rate scales like $\frac{T_{R}^{12}}{\Lambda^{4}m_{g}^{4}}$ (where $\Lambda$ is the effective field theory scale of the modified spin-2 sector), signifying an enhanced yield. This is really exciting, since one can make $m_{g}$ arbitrarily small to increase the freeze-in yield, even for relatively low reheating temperature $\left(T_{R}\right)$~\cite{Bernal2018}. This analysis is, however, flawed and misleading, since the effective theory breaks down at a scale $\Lambda_{5}=(M_{\text{Pl}}m_{g}^{4})^{1/5}$, which is parametrically lower than the scale $\Lambda$ originating from the $2\to2$ massive graviton scattering amplitude, thus rendering the yield calculation wrong beyond the scale $\Lambda_{5}$. This has plagued some of the estimates in the literature for both freeze-out and freeze-in dark matter~\cite{Bernal:2018qlk}. In realistic models of spin-2 particle-mediated dark matter production, such as extra-dimensional scenarios, unitarity is ensured by higher-dimensional coordinate invariance, rendering the model free of such unitarity considerations \cite{Chivukula:2019rij,Chivukula:2019zkt,Chivukula:2020hvi,Chivukula:2023qrt}. 

We therefore invoke extra-dimensions as a model in which gravitational interactions from compactified extra-dimensions can lower the effective quantum gravity scale from $M_{\text{Pl}}$ to the TeV scale, offering us greater freedom for a UV freeze-in without extremely large reheating temperatures. It also offers an intermediate solution to the electroweak hierarchy problem by bringing down the scale of quantum gravity~\cite{Antoniadis1990,ArkaniHamed1998,Randall1999,Randall19991}. For instance, in Randall-Sundrum models~\cite{Randall1999,Randall19991}, the four-dimensional space-time is a slice of the five-dimensional Anti-de Sitter space, bounded by the ultraviolet [UV] and infrared [IR] branes. Such a geometrical setup, in which the dark matter and standard model species are localised on the IR brane, introduces an additional model parameter corresponding to the strength of the extended gravitational interaction between these species~\cite{Randall1999}.

The fact that compactified extra-dimensions respect the higher dimensional gauge/diffeomorphism invariance ensures that even with massive spin-2 mediators, scattering amplitudes grow no faster than $\frac{s}{\Lambda^{2}}$, where $s$ is the centre of mass energy and $\Lambda$ the effective field theory scale of the model (parametrically lower than $M_{\text{Pl}}$)~\cite{Chivukula:2019rij,Chivukula:2019zkt,Chivukula:2020hvi,Chivukula:2023qrt}. A realistic extra-dimensional model also requires the radial model, the radion/dilaton, to be stabilised within the extra-dimensional geometry. The stabilisation process generates a mass for the hitherto massless radion. This is typically achieved via the {\it{Goldberger-Wise}} (GW) mechanism~\cite{Goldberger:1999uk,Goldberger:1999un}, in which a bulk scalar is introduced with non-trivial boundary expectation values. While a naive mass term for a radion reintroduces the bad high energy behaviour for $2\to 2$ scattering of massive gravitons ($\frac{s^{2}}{m_{r}^{2}\Lambda^{2}}$, $m_{r}$ being the mass of the radion), in a properly stabilized extra-dimension, the full tower of GW scalars ensures that amplitudes grow no faster than the usual $\frac{s}{\Lambda^{2}}$, expected from dimensional analysis, power counting and Ward identities~\cite{Chivukula:2021xod,Chivukula:2022tla}. Therefore, although the radion/dilaton portal is naively a simple scalar portal dark matter model, the dynamics are quite non-trivial when considered within a full {\it{UV}}-complete scenario. \looseness=-1

The discussion above is important because any dark matter calculation involves computing scattering amplitudes (or decay rates) to estimate the yield. Several previous calculations within these models have erroneously computed these yields (or misinterpreted the naive scaling in terms of perturbative unitarity), whether in the freeze-out~\cite{Lee:2013bua} or freeze-in mechanisms~\cite{Cai:2021nmk}, whether in models with a single massive spin-2 particle or within a full Kaluza-Klein theory.  \looseness=-1

Within RS models, a comprehensive freeze-out of dark matter (scalars, fermions or vectors) bound to the TeV brane via the Kaluza-Klein spin-$2$ tower and the Goldberger-Wise scalar (spin-$0$) tower has been accurately calculated in~\cite{Chivukula:2024nzt,Chivukula:2025pmk} \footnote{See also Ref.~\cite{deGiorgi:2021xvm}}, for both light~\cite{Chivukula:2024nzt} and heavy radion~\cite{Chivukula:2025pmk}. It was shown that, in the freeze-out scenario, dark matter annihilation was most efficient on resonance with various modes of the spin-2 KK tower or with a heavy ($\mathcal{O}(100-500)$ GeV) radion. For dark matter annihilating predominantly through spin-2 KK resonances, collider bounds push the spin-2 KK masses to the multi-TeV scale, thereby constraining viable dark matter candidates to the TeV range~\cite{Chivukula:2024nzt}. However, one can obtain dark matter masses which satisfy the relic density in the canonical WIMP freeze-out regime, i.e., $100$ GeV-$500$ GeV range for a {\it{heavy}} scalar radion on resonance with the dark matter \footnote{The parameter space is restricted due to internal consistency of the bulk AdS space.}.

Since the effective scale $\Lambda_{\pi}$ is a free parameter in these models, disregarding naturalness arguments, it can span a large range of values below the Planck scale. It is interesting to ask what happens if dark matter, bound to the IR brane, interacts with standard model particles with a much weaker strength (i.e., a large $\Lambda_{\pi}$). In this case, instead of freeze-out, one can populate the observed dark matter abundance via the {\it{freeze-in}} mechanism. We, therefore, assume that there was no/negligible amount of dark matter in the early Universe, and that dark matter, the mediator(s) originating from the extra-dimensional gravitational sector, and the standard model baths are out of thermal (kinetic and chemical) equilibrium with each other.
In this case, there are a few possibilities: 
\begin{enumerate}
    \item If the reheating temperature ($T_{R}$) is high, around the effective KK scale of the extra-dimensional scale $\Lambda_\pi$ (or at least high enough to produce a number of KK modes)
    all KK modes (spin-$2$ and spin-$0$) below $T_{R}$ will be produced on-shell via freeze-in production, and will eventually decay to lower modes, and finally to dark matter. This requires the daunting task of consistently solving coupled Boltzmann equations to determine the relic density. \looseness=-1 
    \item Assuming that the reheating temperature is well below the mass of the $1$st KK graviton, the only on-shell particles that can be produced from the bath are the dark matter and the massless (or massive in phenomenologically realistic models) radial excitation, the radion.
\end{enumerate}

We leave the first case for a detailed follow-up paper while focusing on the second. Even for the second case, if the reheating temperature is above the radion mass, the radion will be produced on-shell, and therefore will lead to a scenario where the radion is first produced on-shell and then subsequently decays with the same coupling strength to the dark matter. One then has to take into account a radion bath and a dark matter bath while solving a Boltzmann equation to accurately compute the relic density. We first consider the simple case where the reheating temperature is low enough that only the dark matter candidate freezes in. We will show that even this simple scenario leads to complications often ignored in the literature.





An important consideration of the FIMP scenario with a massive mediator is that the phase-space distribution function [PSD] of the dark matter species will inherit an average kinetic energy proportional to the mediator's mass. Hence, depending on the reheating temperature, it is entirely possible to end up in a region of parameter space where the present-day PSD of the dark matter species has a significant warm component. This would result in a significant deviation of the matter power spectrum at small scales from that of $\Lambda_{CDM}$, which can be constrained by measurements of the $Ly$-$\alpha$ forest~\cite{Garzilli:2018jqh}. 

In this work, we provide a comprehensive analysis of the gravitationally mediated FIMP scenario by:
\begin{itemize}
    \item Revisiting and expanding on a calculation of the relic abundance for the graviton-mediated production process using full Boltzmann formalism,
    \item Performing a full Boltzmann formalism relic abundance calculation for the production process mediated by the lightest mode in the extra-dimensional theory, the radion.
    \item While FIMPs are notoriously difficult to probe due to the extremely feeble nature of the couplings, we will show that for KeV-GeV scale dark matter, cosmological constraints such as the free-streaming horizon scale, expressed in terms of the $Ly$-$\alpha$ constraints, impose significant restrictions on the parameter space. 
\end{itemize}

The rest of the paper is organised as follows:
\begin{itemize}
 \item In \cref{sec:GravitonMediatedModel}, we present a full Boltzmann-formalism calculation of the present-day relic abundance $\left( \Omega_\Phi h^2 \right)$ for the dark matter species produced via a simple graviton portal. We derive a simple parametric expression for $\Omega_\Phi h^2$ in terms of the model parameters: the dark matter mass $m_\Phi$ and the reheating temperature $T_R$, and use that expression to highlight the phenomenologically relevant region of the parameter space. 
 \item In \cref{sec:RadionPortalFreezeIn}, we analyse the dark matter species produced via a radion portal using the full Boltzmann formalism. We demonstrate the non-thermal nature of the resulting distribution functions and impose constraints on the model's parameter space from the $Ly$-$\alpha$ measurements. In addition, we demonstrate a significant qualitative difference between the parameter spaces of the bosonic (scalar and vector) and fermionic dark matter candidates.
 \item In Appendix \ref{sec:feynrules} we provide the Feynman rules for the graviton- and radion-mediated processes for dark matter production, while in Appendix \ref{sec:colfunc} we expand on the collision functional used for this work.
\end{itemize}

\section{Graviton Mediated Interaction}
\label{sec:GravitonMediatedModel}

\subsection{The Model}
\begin{figure}
    \centering
    \includegraphics[height=5cm]{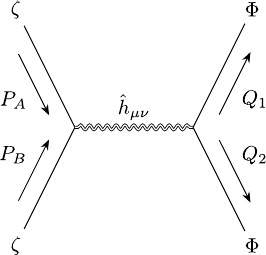}
    \caption{Feynman diagram corresponding to the standard-model species annihilating into the dark-matter species via the graviton portal. Here $\Phi$ denotes the DM species, which can be a scalar, vector, or fermion, denoted by $S$, $V$, and $\chi$, respectively, while $\zeta$ denotes all standard model particles. \label{fig:GravitonnPortalDiag}}
\end{figure}

Since a putative dark matter particle certainly interacts gravitationally with the standard model, we start by considering a simple scenario in which the interaction between the standard model and dark matter thermal baths is mediated by the gravitational field. The particle that carries the gravitational interaction is a massless spin-$2$ excitation (commonly denoted as $h_{\mu \nu}$) defined as a perturbation of the background metric $g_{\mu \nu}$ from the Minkowski space-time $\eta_{\mu \nu}$ as
\begin{equation}
    g_{\mu \nu} = \eta_{\mu \nu} + \frac{2}{M_{\text{PL}}} h_{\mu \nu},
\end{equation}
where $M_{\text{Pl}} = 2.435 \ldots \times 10^{15}$ TeV is the Planck mass. The terms in the action corresponding to an effective interaction between the graviton and the standard model and dark matter sectors in the ``mostly negative'' metric convention are written as 
\begin{equation}
    S \ni \frac{1}{M_{\text{Pl}}} \int_V d^4 x \hat{h}_{\mu \nu} \left( T^{\mu \nu} + \tilde{T}^{\mu \nu} \right), \label{eqn:gravactioninterterms}
\end{equation}
where $T^{\mu \nu}$ is the stress-energy tensor of the standard model and $\tilde{T}^{\mu \nu}$ is the stress-energy tensor of the dark sector, which, for simplicity, we assume to be given by a single massive field $\Phi = \{ S, V, \chi \}$, where $S,V,\chi$ denote scalars, vectors and fermions respectively. The set of Feynman rules and the corresponding matrix elements derived from \cref{eqn:gravactioninterterms} are presented in \cref{sec:GraviFeyn}. Being a higher-dimensional operator, the process of freeze-in is Ultra-Violet (UV), implying that we expect a strong temperature dependence of the collision functional and, eventually, of the relic abundance.

\subsection{Relic Abundance}

\begin{figure}
    \centering
    \includegraphics[height=5.1cm]{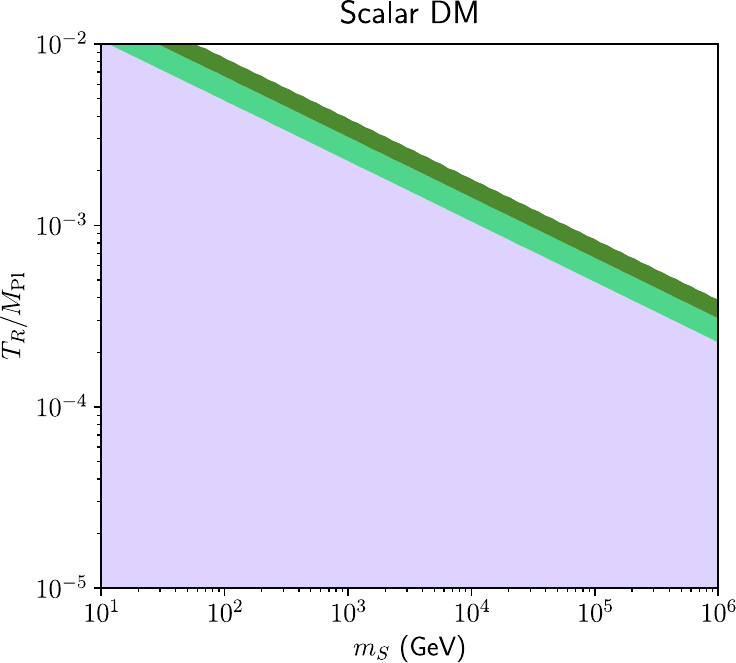}
    \includegraphics[height=5.1cm]{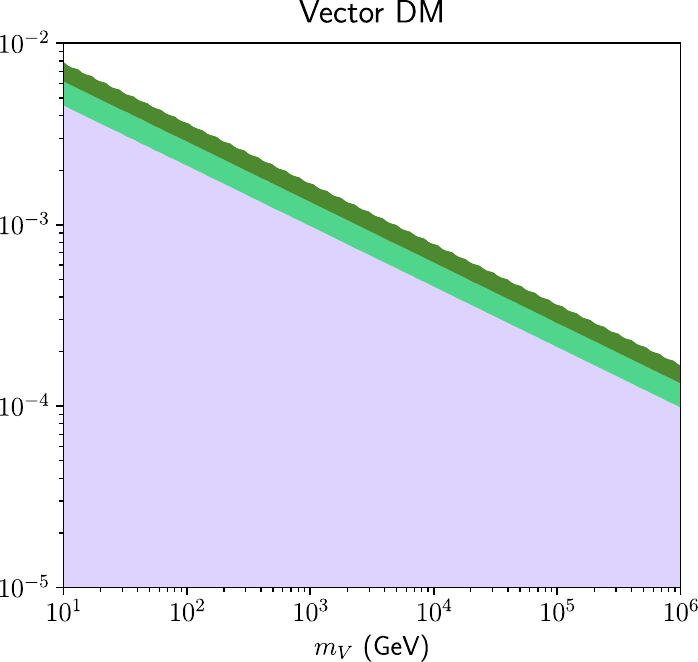}
    \includegraphics[height=5.1cm]{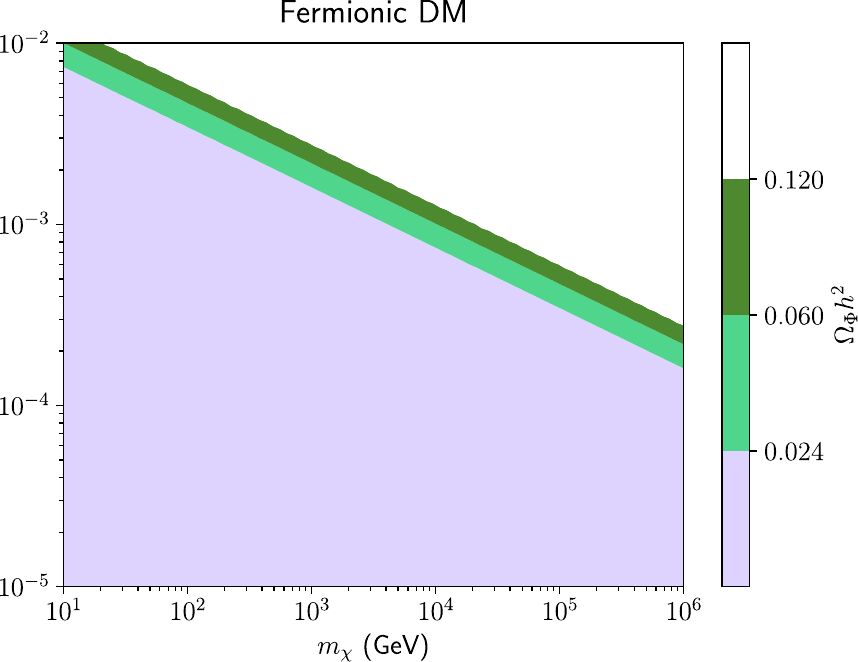}
    \caption{Parameter space of the model of dark matter species $\Phi = \{ S, V, \chi \}$ interacting with the standard model thermal bath through the gravitational force. The plotted region corresponds to $T_R \gg m_\Phi \gg m_\zeta$, where $T_R$ is the reheating temperature, $m_\Phi$ is the mass of the dark matter species, and $m_\zeta$ is the mass of various standard model species. \label{fig:GravitonMediatorParameterSpaceScan}}
\end{figure}

To compute the corresponding abundance, we consider the behaviour of the collision functional in a relevant phenomenological limit. Because the gravitational coupling constant is very small, we expect the relevant region of the parameter space to be $T \gg m_\Phi \gg m_\zeta$, where $m_\Phi$ is the mass of the dark matter, and we use $\zeta$ as shorthand for various standard-model species. The temperature $T$ is a model-dependent quantity corresponding to the temperature of the standard-model bath at the point when dark matter production starts. For this case, assuming instantaneous reheating of the Universe due to a decaying inflaton, it will roughly correspond to the reheating temperature within a factor of a few. \looseness=-1

Furthermore, due to a weakness of the coupling of the gravitational mediator, we can safely neglect the backward contribution to the collision functional. Hence, in this limit, upon the summation over all production channels, we are left with a rather simple expression for the production term in the collision functional
\begin{equation}
    C{\left[ f_\Phi  \right]} \approx \frac{13 \mathcal{S}_\Phi p}{30 \pi^3}   \left( \frac{T}{M_{\text{Pl}}} \right)^4 e^{- \dfrac{p}{T}},\label{eqn:approxcollfuncgrav}
\end{equation}
where $\mathcal{S}_\Phi$ is a numerical prefactor dependent upon the spin of dark matter species $\mathcal{S}_S = 1$, $\mathcal{S}_V = 53/13$, and $\mathcal{S}_\chi = 73/52$. It is worth noting that the values of the prefactor $\mathcal{S}_\Phi$ indicate that the collision functional for the vector dark matter candidate is a factor of $\approx 4$ larger than that of the scalar dark matter candidate and is a factor of $\approx 3$ larger than that of the fermion dark matter candidate; this is analogous to the behaviour observed in the case of the dark matter freeze-out through the spin-$2$ portal in Ref.~\cite{Chivukula2024}. We would also like to point out that \cref{eqn:approxcollfuncgrav} peaks at $p = T$, indicating that the standard model and dark matter baths are in direct thermal contact. Hence, we can conclude that the resulting distribution function of the species $\Phi$ will have a ``temperature'' equivalent to that of the standard model bath. 

We can use \cref{eqn:approxcollfuncgrav} to obtain an insight into the behaviour of the relic density $\Omega_\Phi$. Due to a $\propto T^4$ dependence of the collision functional on the temperature of the standard model bath, it is sufficient to compute the value of the distribution function $f_\Phi$ at some $T = T^\star$, which is proportional to the reheating temperature $T_R$ by some $\order{1}$ factor. This result can then be redshifted to today's temperatures in order to obtain a sufficiently accurate value of the present-day $\Omega_\Phi$. As we are concerned with the production of species $\Phi$ in the radiation-dominated universe, we can link the temperature of the standard model bath $T$ to cosmic time $t$ as per \cref{eqn:timetemperaturelink}, allowing us to write down the relation
\begin{equation}
    \frac{dt}{dT} = - \frac{\sqrt{90} M_{\text{Pl}}}{\pi \sqrt{g^*{\left( T \right)}}} \frac{1}{T^3}, \label{eqn:timetemplink2}
\end{equation}
where $g^*{\left( x \right)}$ is the number of open degrees of freedom of the standard model bath. Parametrising the momentum of species $\Phi$ $p$ in terms of present-day momentum $q$ as $p = q / a{\left( T \right)}$ where $a{\left( T \right)}$ is the scale factor at the temperature of the standard model bath $T$ and using \cref{eqn:timetemplink2}, we can express the transport equation \cref{eqn:BoltzmanEqn} as 
\begin{equation}
    \frac{\partial f_\Phi}{\partial T} = - \frac{\sqrt{90} M_{\text{Pl}}}{\pi \sqrt{g^*{\left( T \right)}}} \frac{C{\left[ f_\Phi \right]}}{T^3}, \label{eqn:BoltzTempApprox}
\end{equation}
where $f_\Phi$ is a function of today's momentum $q$ and temperature of standard model bath $T$. Introducing our approximation \cref{eqn:approxcollfuncgrav} into \cref{eqn:BoltzTempApprox} and performing integration over the temperature from $T_R$ down to $T^\star$ we get the following integral expression for the distribution function $f_\Phi$ 
\begin{equation}
    f_\Phi{\left( q, T \right)} = \frac{13 \sqrt{90} \mathcal{S}_\Phi}{30 \pi^4 M_{\text{Pl}}^3} q \int_{T^\star}^{T_R} dT \frac{T}{a{\left( T \right)} \sqrt{g^*{\left( T \right)}}}  e^{- q/ a{\left( T \right)} T}. \label{eqn:approxfinteg}
\end{equation}
As both $T_R$ and $T^\star$ are much larger than the masses of the standard model species, we can safely assume that $g^*{\left( T \right)} = g^*{\left( T_R \right)}$ $\forall$ $T \in \left[ T^\star, T_R \right]$. Furthermore, as we are in the radiation-dominated universe, the product $a{\left( T \right)} T$ is constant, allowing us to also assume that $a{\left( T \right)} T = a{\left( T_R \right)} T_R$ $\forall$ $T \in \left[ T^\star, T_R \right]$. Thus, we can factor constants out of the integral \cref{eqn:approxfinteg}, resulting in 
\begin{align}
    f_\Phi{\left( q, T \right)} &= \frac{13 \sqrt{90} \mathcal{S}_\Phi}{30 \pi^4 M_{\text{Pl}}^3  \sqrt{g^*{\left( T_R \right)}} }  \frac{q}{a{\left( T_R \right)} T_R} e^{- q/ a{\left( T_R \right)} T_R} \int_{T^\star}^{T_R} dT T^2 \\ &\approx \frac{13 \mathcal{S}_\Phi }{\sqrt{90 g^*{\left( T_R \right)}} \pi^4  }  \frac{q}{a{\left( T_R \right)} T_R} e^{- q/ a{\left( T_R \right)} T_R} \left( \frac{T_R}{M_{\text{Pl}}} \right)^3, 
\end{align}
where, as $T_R > T^\star$, we have assume that $T_R^3 \gg \left( T^\star \right)^3$. Lastly, we impose the conservation of entropy 
\begin{equation}
    T_R a{\left( T_R \right)} = \left( \frac{g^*{\left( T_0 \right)}}{g^*{\left( T_R \right)}} \right)^{1/3} T_0,
\end{equation}
where $T_0$ is the present-day temperature of the standard model bath, allowing us to obtain the following approximation for the present-day distribution function $f_\Phi$
\begin{equation}
    f_\Phi{\left( q, T_0 \right)} \approx \frac{13 \mathcal{S}_\Phi}{\sqrt{90 g^*{\left( T_R \right)}} \pi^4  } \left( \frac{g^*{\left(T_R  \right)}}{g^*{\left(  T_0  \right)}} \right)^{1/3} \left( \frac{T_R}{M_{\text{Pl}}} \right)^3 \frac{q}{T_0} e^{- \left( \dfrac{g^*{\left(T_R  \right)}}{g^*{\left(  T_0  \right)}} \right)^{1/3} \dfrac{q}{T_0} }. \label{eqn:fPhiApproToday}
\end{equation}
At this stage, we have everything ready to perform the integration over present-day momentum $q$ in order to obtain the relic density $\Omega_\Phi$ as per \cref{eqn:OmegaDefinition}. To simplify the result even further, we can perform another approximation, as due to the species $\Phi$ being heavy, it is safe to assume that the present-day distribution function $f_\Phi$ is non-relativistic; hence, we can write the energy $\sqrt{p^2 + m_\Phi^2}$ appearing in \cref{eqn:OmegaDefinition} as simply $m_\Phi$. Therefore, we obtain the following expression for the relic density
\begin{equation}
    \Omega_\Phi h^2 \approx \frac{39 \mathcal{S}_\Phi g^*_\Phi m_\Phi }{\sqrt{90 g^*{\left( T_R \right)}} \pi^6  }  \frac{g^*{\left(T_0  \right)}}{g^*{\left(  T_R  \right)}}   \left( \frac{T_R}{M_{\text{Pl}}} \right)^3 \frac{T_0^3 h^2}{3 M_{\text{Pl}}^2 H_0^2}, \label{eqn:OmegaAnalyticForm}
\end{equation}
where $g^*_\Phi$ is the number of degrees of freedom carried by the DM species $\Phi$ and $H_0$ is the present-day Hubble constant. In this work, we find that the parametric expression
\begin{equation}
    \Omega_\Phi h^2 \approx \left( 2.1 \times 10^3 \right) \mathcal{S}_\Phi g^*_\Phi \left( \frac{m_\Phi}{1 \text{ GeV}} \right) \left( \frac{T_R}{M_{\text{Pl}}} \right)^3,  \label{eqn:OmegaParametricForm}
\end{equation}
agrees extremely well with our direct numerical solution of \cref{eqn:TransportEqnFull} in the region of parameter space where $T_R \gg m_\Phi \gg m_\zeta$. \cref{eqn:OmegaParametricForm} also agrees extremely well with our analytic approximation \cref{eqn:OmegaAnalyticForm}. The plot of the scan over this region of the parameter space for various dark matter candidates is presented in \cref{fig:GravitonMediatorParameterSpaceScan}.

This analysis demonstrates a large region of parameter space that can account for the observed dark matter relic density across a larger range of dark matter masses $m_\Phi \simeq 10$ GeV $-$ $1000$ TeV for large reheating temperatures $T_R \simeq$ $10^{-4} M_{\text{Pl}} - 10^{-2} M_{\text{Pl}}$ where, due to the weakness of the gravitational interaction, any relevant cross-sections fall well below current dark matter direct detection and collider bounds. In addition, due to the dark matter masses required to saturate the relic density with reheating temperatures under $1\%$ of the Planck mass being well above the scales of a few keV, no Lyman-$\alpha$ constraints can be applied as the effects of this dark matter candidate on the present-day matter power spectrum are indistinguishable from those of cold dark matter. The viable region of the parameter space also extends to sub-GeV dark matter masses. However, saturating the relic density in this region would require reheating temperatures greater than $1\%$ of the Planck mass, where some of the assumptions required for the effective field theory \cref{eqn:gravactioninterterms} may no longer hold. The viable region of parameter space also extends to the above-$1000$ TeV dark matter candidate mass range, where $m_\Phi$ of an order of $T_R$ is required to saturate the relic density and our simple expression \cref{eqn:OmegaParametricForm} no longer holds. \looseness=-1

At this point, a few caveats about the scenario are worth noting, especially regarding potential bounds on the reheating temperature. The reheating temperature can be constrained using analysis of Planck satellite data for a variety of inflationary models. The Planck data is compatible with Higgs inflation with a reheating temperature of $10^{13}$ GeV, as well as with $R^{2}$ inflation, which predicts a reheating temperature of $10^{9}$ GeV~\cite{Planck:2018jri}. Further analysis shows that assuming instantaneous reheating, $\alpha$-attractor models with a reheating temperature of $10^{15}$ GeV are allowed within a confidence level of $68$ $\%$~\cite{Ellis:2025zrf}. We remain agnostic about the reheating temperature scale here, but remind the reader that, in this scenario, the only constraints on the model will arise from cosmological data for a given inflationary model. Note that if we take the Higgs inflation upper bound of $10^{13}$ GeV, for DM masses of $1$ TeV, the relic abundance is found to be around $10^{-10}$, which is tiny. For this model, to obtain a reasonable relic fraction of order $10^{-2}$, we would need a DM mass on the order of $10^{10}$ GeV. Model building with such large dark matter can be challenging, although models such as WIMPzillas predict ultra-heavy dark matter candidates \cite{Kolb:1998ki}. To conclude this brief section, we simply remind the reader that this process is always present due to the universal nature of gravity, and depending on the scale of the reheating temperature and the dark matter mass, this can indeed provide an $\mathcal{O}{\left(1 \right)}$ fraction of the observed relic abundance.

\section{Freeze-in from a Radion Mediated Interaction in Randall-Sundrum model}
\label{sec:RadionPortalFreezeIn}

Since we are interested in freeze-in processes mediated by the gravitational sector, one way to circumvent the high reheating temperature is to lower the interaction scale, as the resulting relic abundance is inversely proportional to it. The most straightforward way to achieve this is to consider extra-dimensional set-ups. In this particular instance, we will consider the Randall-Sundrum (RS) model \cite{Randall:1999ee,Randall:1999vf} as a prototypical example of an extra-dimensional model, which provides an intermediate solution to the electroweak hierarchy problem. The basic set-up is to define the fundamental theory on a $5$-dimensional manifold with an Anti-de-Sitter bulk space, and two branes (Planck (UV) and TeV (IR)) as flat Minkowski slices. The warped bulk generates an exponential hierarchy of scales between the Planck and the TeV brane, bringing the scale of quantum gravity on the TeV brane from $M_{Pl}$ to the TeV scale. In the subsequent subsection, we briefly describe the model setup and the assumptions pertinent to this work.

\subsection{The Model}
Following our previous works \cite{Chivukula:2019rij,Chivukula:2020hvi,Chivukula:2021xod,Chivukula:2022tla}, the $5$-dimensional action in Unitary gauge for the gravitational sector of the stabilised RS model is written as, 
\begin{equation}
    S = \int_V d^4 x~ d y \left( \mathcal{L}_{\text{EH}} + \mathcal{L}_{\phi \phi} + \mathcal{L}_{\text{pot}} + \mathcal{L}_{\text{GHY}} + \Delta \mathcal{L} \right),
    \label{eq:stablag}
\end{equation}
The first three terms $\mathcal{L}_{\text{EH}},\mathcal{L}_{\phi \phi}$ and $\mathcal{L}_{\text{pot}}$ represent the 5D Einstein-Hilbert, the bulk scalar kinetic term and the scalar potential which stabilizes the bulk of the extra-dimension with non-trivial boundary terms on the brane. $\mathcal{L}_{\text{GHY}}$ is the Gibbons-Hawking-York term \cite{Gibbons:1976ue,York:1972sj} that ensures the action yields a well-posed variational problem by accounting for boundary terms. Finally, the total derivative term $\Delta\mathcal{L}$ is added to yield canonical kinetic terms without any loss of generality. The details of each term can be found in \cite{Chivukula:2022tla}.

In unitary gauge, the $5$D metric for the stabilized AdS background can be parametrized as~\cite{Chivukula:2021xod,Chivukula:2022tla,Charmousis:1999rg} 
\begin{equation}
    G_{M N} = \begin{pmatrix} w g_{\mu \nu} & 0 \\ 0 & -v^2 \end{pmatrix} \text{ and } G^{M N} = \begin{pmatrix} g^{\mu \nu} / w & 0 \\ 0 & - 1 / v^2 \end{pmatrix},\, 
    \label{eqn:metricparametrisation}
\end{equation}
with co-ordinates $x^M=(x^\mu,y)$, where $y \in (-\pi r_c,+\pi r_c]$ parametrizes an orbifolded extra-dimension. We define $\varphi \equiv y/r_c$, for notational convenience.
The tensor fluctuations over the flat $4$D background $\eta_{\mu\nu}$ (in the``mostly-minus" convention) are parametrized as 
\begin{align}
    g_{\mu\nu}\equiv \eta_{\mu\nu} + \kappa \hat{h}_{\mu\nu}(x_{\mu},y)~,  \label{eq:metricparam1}
\end{align}
where $\hat{h}_{\mu\nu}(x,y)$ parametrizes metric fluctuations. The quantities $w$ and $v$ are defined as, 
\begin{align}
    w &=  \exp\left[-2\left(A(y)+\frac{ e^{2A(y)}}{2\sqrt{6}} \kappa\,\hat{r}(x,y)\right)\right] \nonumber \\
    v &= \left(1 + \frac{ e^{2A(y)}}{\sqrt{6}} \kappa\,\hat{r}(x,y)\right)~,
    \label{eq:metricparam2}
\end{align}
where the $A(y)$ specifies the background geometry. In the unstabilized RS model, the warp factor is simply $A(y) = -k|y|$, which yields a bulk Anti-de Sitter (AdS) geometry. In the stabilised model, the function $A(y)$ has two contributions, one from the unstabilized RS model, and a back-reaction term that originates from the stabilisation potential (encoded in $\mathcal{L}_{pot}$), driven by a bulk scalar field, and has an opposite sign compared to the ADS solution. With the above parameterisation of the metric, the tensor and scalar fluctuations can be written, respectively, as \cite{Chivukula:2022tla}, 

\begin{align}
    \hat{h}_{\mu\nu}(x,y) &=  \dfrac{1}{\sqrt{ r_{c}}} \sum_{n=0}^{+\infty} \hat{h}^{(n)}_{\mu\nu}(x)\, \psi_{n}(\varphi)~,
    \label{eq:spin2mode-equation} \\
    \hat{r}(x,y) &= \dfrac{1}{\sqrt{ r_{c}}} \sum_{i=0}^{+\infty} \hat{r}^{(i)}(x)\, \gamma_{i}(\varphi)~.
    \label{eq:spin0-mode-expansion}
\end{align}

\begin{figure}
    \centering
    \includegraphics[height=5cm]{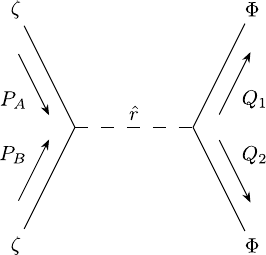}
    \caption{Feynman diagram corresponding to the standard model species annihilating into the dark matter species through the radion portal. Here $\Phi$ corresponds to the DM species, which can be either a scalar, vector, or a fermion, denoted by $S$, $V$, and $\chi$ respectively, while $\zeta$ denotes all kinds of the standard model particles. \label{fig:RadionPortalDiag}}
\end{figure}

As discussed in the previous section of this work, in the simplest gravitationally coupled model, the reheating temperatures required to saturate the relic density are very large. Given that the collider bounds on the spin-2 KK sector are on the order of a few TeV and we are interested in low reheating-temperature scenarios, we consider the radion to be the mode through which resonant production of frozen-in dark matter occurs. Thus, for the purpose of this work, only the dynamics of scalar fluctuations is relevant.

The terms in the action correspond to an effective interaction between the radion and the standard model and dark matter sectors are written as
\begin{equation}
 S \ni - \frac{1}{\sqrt{6} \Lambda_\pi} \int_V d^4 x \hat{r} \left( T_\mu^\mu + \tilde{T}_\mu^\mu \right), \label{eqn:actioninterterms}
\end{equation}
where $\Lambda_\pi$ is the cut-off scale of the theory, which is treated as a free parameter in this work, $\hat{r}$ is the radion field, $T_\mu^\mu$ is the trace of the stress-energy tensor of the standard model, $\tilde{T}_\mu^\mu$ is the trace of the stress-energy tensor of the dark sector, and $\sqrt{6}$ is a conventional normalisation factor. We refer the reader to the discussion in Ref.~\cite{Chivukula2024}, which demonstrates how the interaction terms of the form \cref{eqn:actioninterterms} naturally arise in the extra-dimensional scenario. A mass term for the radion can be obtained via the Goldberger-Wise mechanism by introducing a bulk scalar field with a non-trivial vacuum expectation value as briefly discussed earlier.

As discussed in \cite{Chivukula:2025pmk}, an important peculiarity of this model is that there is an upper bound on the radion mass arising from the eventual switch in the sign of the background curvature in the vicinity of the Planck brane from anti-de Sitter to de Sitter. This issue is, however, circumvented for the radion masses considered in this work because the coupling scale in the FIMP scenario is much larger than the typical WIMP scale of $\order{10 \text{TeV}}$. 

\subsection{Relic Abundance}

The cosmoligical relic abundance is obtained via SM particles interacting with dark matter via the radion. The relevant matrix elements are listed in Appendix \ref{sec:feynrules} .
 We observe that the matrix elements (for example in\cref{eqn:SmatrixHHChiChi,eqn:SmatrixZZChiChi,eqn:SmatrixWWChiChi,eqn:SmatrixFFChiChi}) are independent of the Mandelstam variable $t$ and contain the $\delta$-function that fixes the Mandelstam variable $s$ to the square of the radion mass. Therefore, the integral over $t$ appearing in \cref{eqn:collisionfuncdef} can be performed trivially with the help of \cref{eqn:tminmaxdef}
\begin{equation}
    \int_{t^{\text{min}}{\left( s \right)}}^{t^{\text{max}}{\left( s \right)}} dt \delta{\left( s - m_r^2 \right)} = \int_{t^{\text{min}}{\left( m_r^2 \right)}}^{t^{\text{max}}{\left( m_r^2 \right)}} dt \delta{\left( s - m_r^2 \right)}= m_r \sqrt{m_r^2 - 4 m_\Phi^2} \sqrt{1 - \frac{4 m_\zeta^2}{m_r^2}} \delta{\left( s - m_r^2 \right)}. \label{eqn:tintegral}
\end{equation}
Hence, upon substituting \cref{eqn:SmatrixHHChiChi,eqn:SmatrixZZChiChi,eqn:SmatrixWWChiChi,eqn:SmatrixFFChiChi} into \cref{eqn:collisionfuncdef} and subsequent integration over Mandelstam variables with the help of \cref{eqn:tintegral} and summation over all standard model species $\zeta$\footnote{Technically, summation over the branching ratios appearing in \cref{eqn:SmatrixHHChiChi,eqn:SmatrixZZChiChi,eqn:SmatrixWWChiChi,eqn:SmatrixFFChiChi} will yield $1 - \beta{\left( r \rightarrow \Phi \Phi \right)}$. However, we assume that the branching ratio into the dark sector $\beta{\left( r \rightarrow \Phi \Phi \right)}$ is negligible due to the radion having many more decay channels into the standard model species. This assumption is valid for reheating temperatures below the electroweak symmetry-breaking scale.}, we are left with a rather simple expression
\begin{multline}
C{\left[ f_\Phi \right]} = \frac{1}{96 \pi g^*_\Phi E p \Lambda_\pi^2} \upsilon_\Phi{\left( m_r, m_\Phi \right)} \times \\ \left[ T e^{-E/ T} \left( e^{-E^{\text{min}}{\left( m_r^2 \right)} / T} - e^{-E^{\text{max}}{\left( m_r^2 \right)} / T}  \right) - f_\Phi \left( F{\left( E^{\text{max}}{\left( m_r^2 \right)}\right)} - F{\left( E^{\text{min}}{\left( m_r^2 \right)}\right)} \right) \right], \label{eqn:collisionfuncradion}
\end{multline}
where the first term in the square brackets corresponds to the contribution from the forward process (annihilation of the standard model species into the dark matter species) and the second term in the square brackets corresponds to the backwards process (annihilation of the dark matter species into the standard model species).
An additional insight into the behaviour of this process can be obtained by considering a phenomenologically relevant limit in which the mass of the dark matter candidate is much smaller than the radion mass, and production occurs at temperatures well below the radion mass. Taking the $m_\Phi \ll m_r$ limit of \cref{eqn:collisionfuncradion} at the point where the production of species $\Phi$ starts ($f_\Phi = 0$) we get
\begin{equation}
    C{\left[ f_\Phi = 0 \right]} \approx \frac{T}{96 \pi g^*_\Phi}  \frac{\bar{\upsilon}_\Phi{\left( m_r, m_\Phi \right)}}{p^2 \Lambda_\pi^2}  e^{ - \left( \dfrac{p}{T} + \dfrac{m_r^2}{4 p T} \right)}, \label{eqn:ExpansionOfCollFunc}
\end{equation}
where $\bar{\upsilon}_\Phi{\left( m_r, m_\Phi \right)}$ is the large radion mass expansion of the interaction-dependent factors \cref{eqn:upsilonS,eqn:upsilonV,eqn:upsilonChi} given by
\begin{align}
    \bar{\upsilon}_S{\left( m_r, m_S \right)} &= \bar{\upsilon}_V{\left( m_r, m_V \right)} = m_r^4\label{eqn:fac_scalar_vector_fermion}, \\
    \bar{\upsilon}_\chi{\left( m_r, m_\chi \right)} &= 2 m_\chi^2 m_r^2.
\end{align}
\cref{eqn:ExpansionOfCollFunc} is a rather simple function with a peak at 
\begin{equation}
    p = \frac{m_r}{2} \left( \sqrt{ 1 + \frac{4 T^2}{m_r^2}} - \frac{2 T}{m_r} \right). \label{eqn:phenolimitpeak}
\end{equation}
\cref{eqn:phenolimitpeak} indicates that for the case of $T \ll m_r$, the peak of the collision functional and, by consequence, the resulting distribution function of species $\Phi$ is given entirely in terms of the radion mass $m_r$. In other words, we are able to produce dark matter species with a ``temperature'' that depends on the mediator mass rather than on the temperature of the standard model bath. 

\subsection{Phase Space Distribution (PSD) of dark matter}
\label{sec:PSD}

The presence of a ``non-cold" component in the total dark matter relic manifests itself as a suppression of the matter power spectrum at small scales (large wave-numbers)~\cite{Diamanti:2017xfo,Murgia:2017lwo}. This is because non-cold dark matter (NCDM) has a non-equilibrium present-day velocity distribution, unlike cold dark matter, which obeys a Maxwellian distribution. As a result, NCDM free-streams to greater lengths, washing out structures at scales below its free-streaming length, and this shows up as a suppression of the matter power spectrum relative to CDM-like behaviour. The velocity spread of NCDM candidates is often modelled using production mechanisms that yield a thermal (Maxwell-Boltzmann-like) distribution function~\cite{DEramo:2020gpr}, with such species commonly referred to as ``warm'' dark matter (WDM). However, the mass of thermally produced warm dark matter is tightly constrained by the cosmological Lyman-$\alpha$~\cite{Viel:2013fqw,Irsic:2017ixq} constraints. In our model, the production mechanism yields a unique skewed-Gaussian (non-thermal) distribution function with a sharp momentum peak rather than a broad thermal spread. 
We will show how the distribution function changes with the tunable parameters of our model: the mass of the dark matter candidates $\left( m_\Phi \right)$ and the ratio of the radion mass to the reheating temperature $\left( m_r / T_R \right)$. Furthermore, it will provide insights into the behaviour of the number density $\left( n_\chi \right)$ as a function of the cut-off scale of the theory $\left(\Lambda_\pi\right)$.

\begin{figure}
    \centering
    \includegraphics[height=5.1cm]{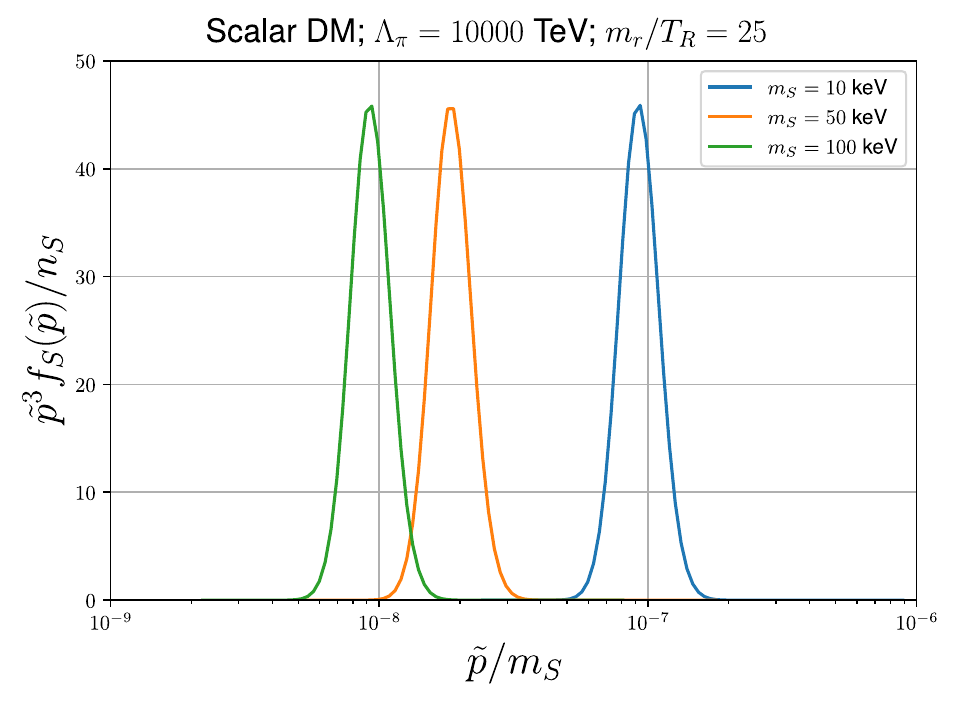}
    \includegraphics[height=5.1cm]{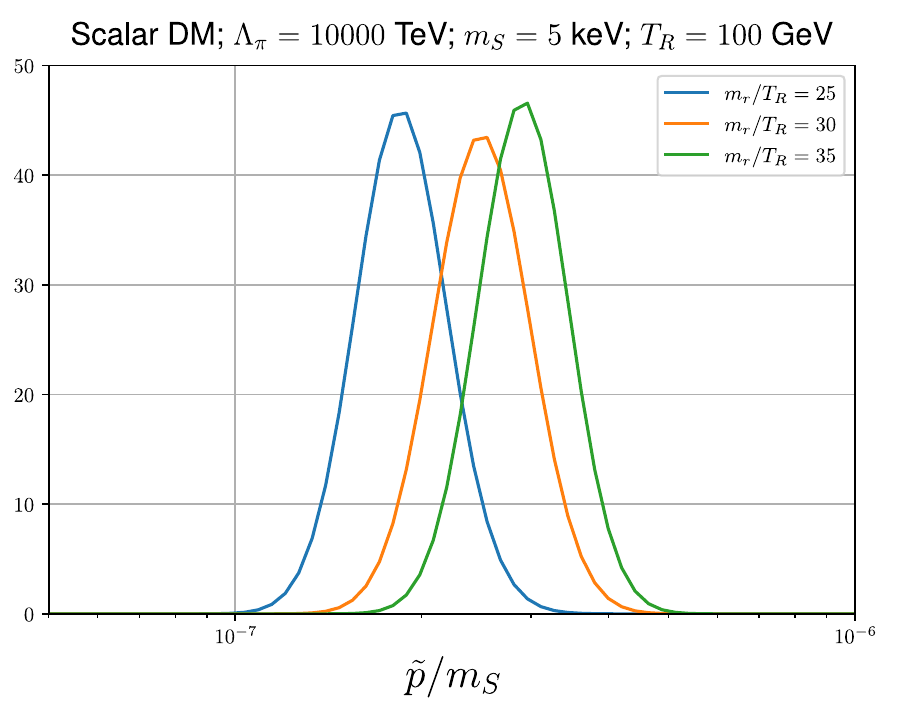}
    \caption{Normalised distribution function for different masses of scalar dark matter species (Left Panel); Normalised distribution function for different values of the $m_r / T_R$ ratio (Right Panel). Distribution functions are plotted as a function of the present-day momentum $\tilde{p}$. The title card of the plots lists fixed model parameters. \label{fig:PSDshape}}
\end{figure}

In the left panel of \cref{fig:PSDshape}, we show how the present-day distribution functions vary with the dark matter mass for fixed values of $m_r/T_R$ and $\Lambda_\pi$. We note that increasing the mass of the dark matter species shifts the distribution function to the left, indicating that a heavier DM particle behaves as a relatively colder relic with a lower peak velocity. This is consistent with \cref{eqn:phenolimitpeak}, as in the regime where $m_r / T_R \gg 1$, the momentum of the peak of the distribution function at the point where the production starts is proportional to the radion mass. This, however, leads to a lower present-day velocity, which is proportional to the ratio of present-day momentum to the dark matter mass, with the radion mass held fixed. On the right-hand side of \cref{fig:PSDshape}, we vary the $m_r / T_R$ ratio for a fixed DM mass and cut-off scale. It is important to note that varying the $m_r/T_R$ ratio involves changing both the radion mass and the reheating temperature. Hence, for simplicity, we fix the reheating temperature to $100$ GeV while keeping the DM mass at $5$ keV. Thus, the ratio $m_r / T_R$ is varied by solely changing the radion mass. It is clear that the peak of the distribution shifts to the right for higher radion masses, which is once again consistent with \cref{eqn:phenolimitpeak}, since the momentum of the dark matter species is proportional to the radion mass. 

\begin{figure}
    \centering
    \includegraphics[height=4.3cm]{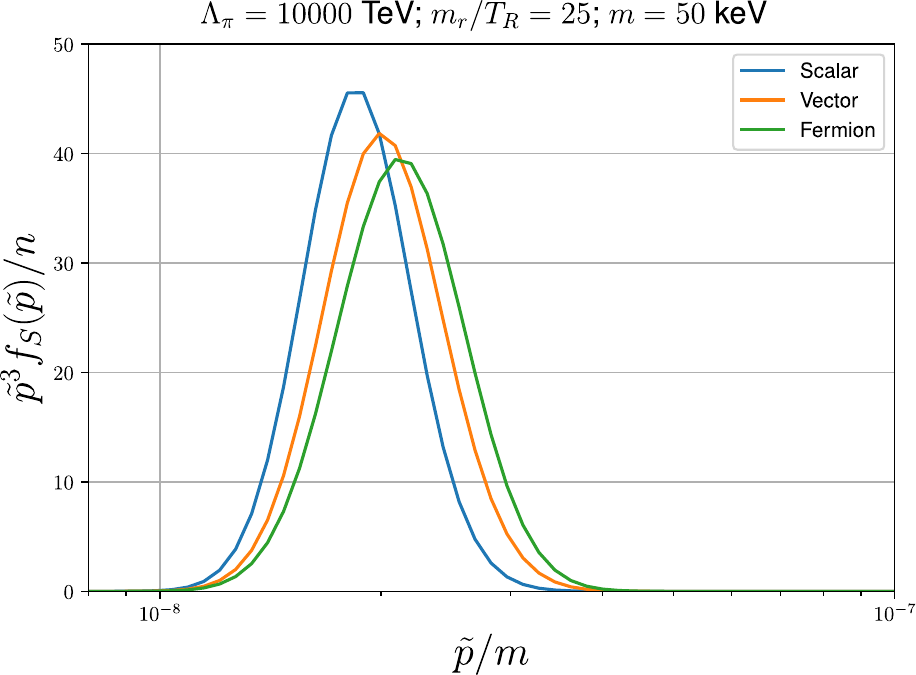}
    \includegraphics[height=4.3cm]{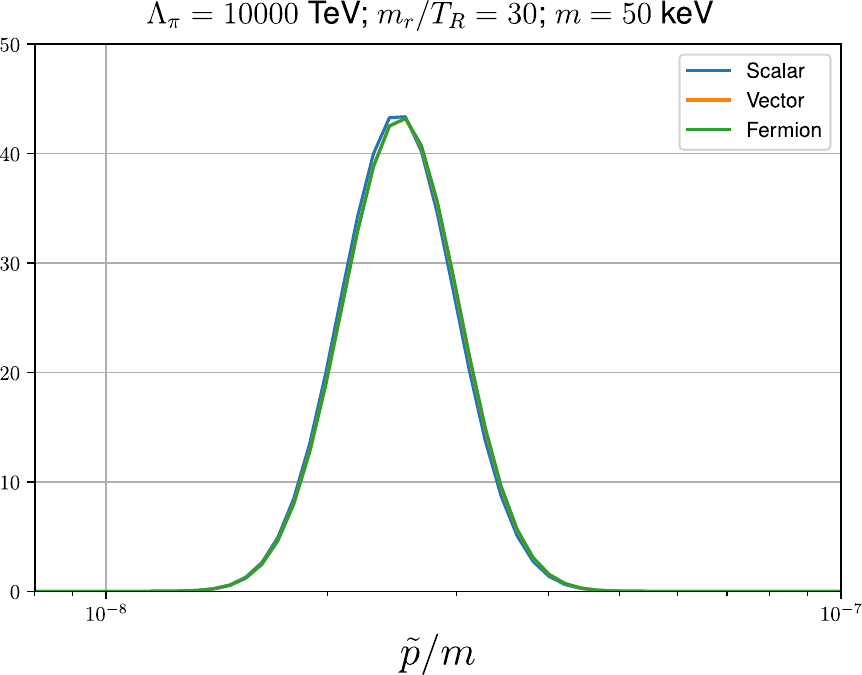}
    \includegraphics[height=4.3cm]{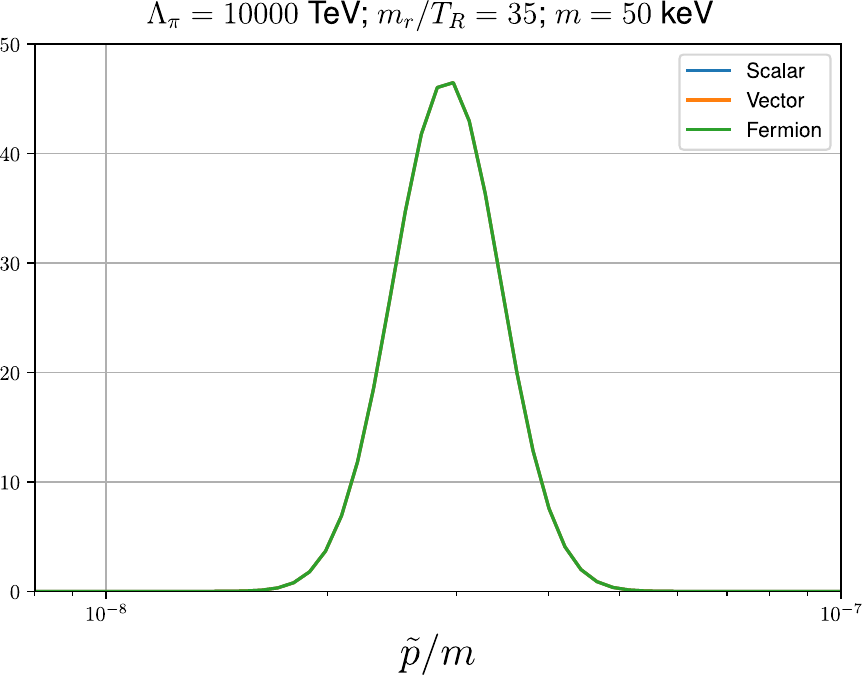}
    \caption{Normalised distribution functions for different spins of the dark matter species with the reheating temperature fixed at $T_R=100$ GeV. Distribution functions are plotted as a function of the present-day momentum $\tilde{p}$. The title card of the plots lists fixed model parameters.\label{fig:PSDspin} }
\end{figure}

Next, we examine how the distribution function varies across the different spins of the DM species for a fixed set of model parameters. This behaviour is illustrated in \cref{fig:PSDspin}. It is worth noting that scalar, vector, and fermion distribution functions are qualitatively very similar once normalised by the final number density in the regime where the radion mass is sufficiently large. This occurs due to the fact that the backward contribution to the collision functional is heavily suppressed and the transport equation is reduced to a linear differential equation with the spin-dependent factor $\upsilon_\Phi{\left( m_r, m_\Phi \right)}$ only affecting the normalisation of the solution. We would like to point out that, as shown on \cref{fig:PSDspin}, this is not the case for some values of the radion mass, for example $m_r / T_R = 25$ on the left plot, in which case the backward term cannot be ignored, leading to a different qualitative behaviour between dark matter candidates of different spins \footnote{We have checked that the freeze-in criteria still holds, i.e the backward reaction is not large enough to cause .}.


\begin{figure}[t]
    \centering
    \includegraphics[width=0.6
    \linewidth]{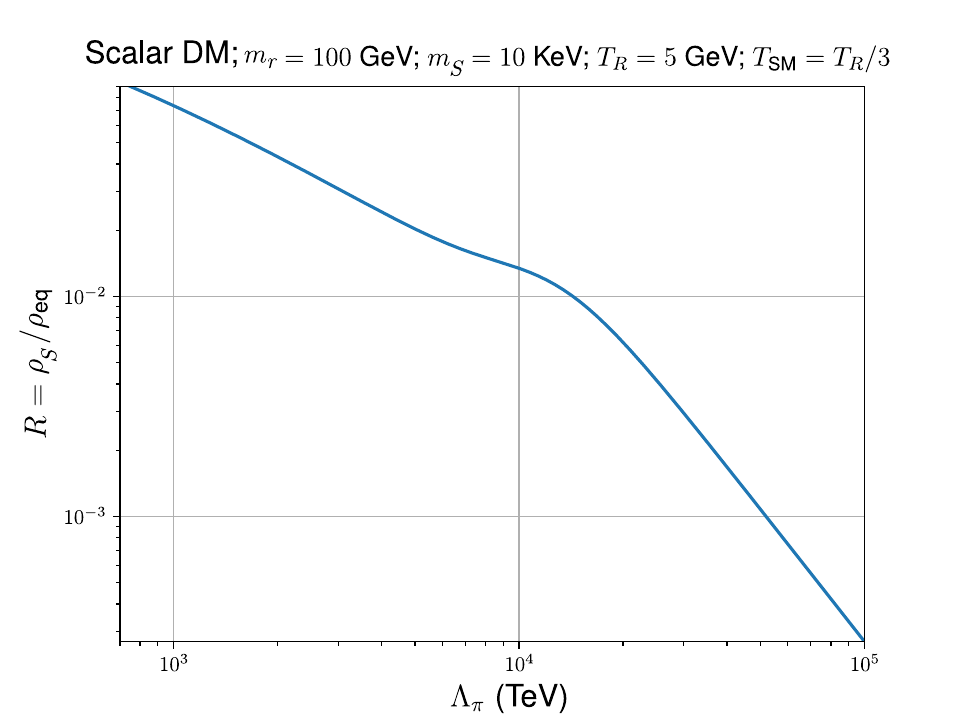}
    \caption{Plot of the ratio of the relic density of the scalar dark matter candidate produced via the radion portal freeze-in to the equilibrium relic density at a given standard-model bath temperature ($T_{SM}$) as a function of the cut-off scale $\Lambda_\pi$. The title card of the plot indicates the model parameters used and the corresponding value of $T_{SM}$.  \label{fig:Rhoratio}}
\end{figure}

Lastly, we focus on the behaviour of the distribution function for different values of the coupling strength, as defined by the cut-off scale $\Lambda_\pi$. To understand this, we start by considering whether or not the dark matter and standard model baths reach equilibrium as the coupling strength changes. We can consider the behaviour of the energy density
\begin{equation}
    \rho_\chi{\left( T_{\text{SM}} \right)} = \int_0^\infty \frac{d^3 p}{\left( 2 \pi \right)^3} \sqrt{p^2 + m_\chi^2} f_\chi{\left( p, T_{\text{SM}} \right)},
\end{equation}
for some temperature of the standard model bath $T_{\text{SM}}$ relative to the equilibrium energy density given by
\begin{equation}
    \rho^{\left( \text{eq} \right)}_\chi{\left( T_{\text{SM}} \right)} = \int_0^\infty \frac{d^3 p}{\left( 2 \pi \right)^3} \sqrt{p^2 + m_\chi^2} e^{- \sqrt{p^2 + m_\chi^2} / T_{\text{SM}}},
\end{equation}
where we have set the distribution function to the Maxwell-Boltzmann one with the corresponding temperature of the standard model bath. It is convenient to look at the behaviour of the ratio
\begin{equation}
    R{\left( T_{\text{SM}} \right)} = \frac{\rho_{\chi}{\left( T_{\text{SM}} \right)}}{\rho_\chi^{\left( \text{eq} \right)}{\left( T_{\text{SM}} \right)}}, \label{eqn:RRatioDef}
\end{equation}
with a change in $\Lambda_\pi$. This behaviour for the scalar dark matter candidate is illustrated in \cref{fig:Rhoratio}, which shows that increasing the cut-off scale (decreasing the coupling strength) moves the system further away from equilibrium. Thus, the ratio of the energy densities decreases. For a freeze-in scenario, we want the energy density of our DM species to be much smaller than its equilibrium energy density at the time of decoupling. Thus, we can make an important observation from \cref{fig:Rhoratio}, as at around $\Lambda_{\pi}= 10000$ TeV, the energy density at the time of decoupling is $\sim 1 \%$ of the equilibrium energy density. This gives us a reasonable choice for the cut-off scale $\Lambda_{\pi}$ for this work and the parameter scans of \cref{fig:RadionMediatorParameterSpaceScan}. Note that the overall qualitative behaviour is identical for the dark matter species of a higher spin.

Moreover, we would like to draw the reader's attention to the fact that the qualitative behaviour illustrated on \cref{fig:Rhoratio} becomes significantly more complicated than the expected power-law scaling form $\abs{\mathcal{M}}^2 \propto 1/\Lambda_\pi^2$ in the regime where the values of $\Lambda_\pi$ are sufficiently small ($\Lambda_\pi \lesssim 2 \times 10^4$ TeV for the set of model parameters used to construct \cref{fig:Rhoratio}) to allow for a non-negligible contribution from the backwards term in \cref{eqn:collisionfuncdef}. Thus, illustrating that even at energy densities of dark matter species as low as $\order{1\%}$ of the equilibrium density, the contribution from the backwards term in \cref{eqn:collisionfuncdef} can be considerable.

\subsection{$Ly$-$\alpha$ Constraints}
\label{sec:Lymanconst}

As in this model, we have control over the mediator coupling via the cut-off scale $\Lambda_\pi$, we must consider experimental constraints on the parameter space arising from the $Ly$-$\alpha$ measurements. The $Ly$-$\alpha$ forest is a set of absorption lines produced by the Intervening Intergalactic Medium (IGM) containing neutral hydrogen in the spectra of distant bright quasars along different lines of sight. Hence, it is a useful probe to constrain the matter power spectrum $P{\left( k \right)}$ on scales $0.5 \text{Mpc}/h < \lambda < 100 \text{Mpc}/h $ \cite{Murgia:2017lwo}. We briefly describe the process in Appendix \ref{sec:lya}.

We have calculated $P(k)$ numerically within the linear perturbation of structure formation and compared our results with the well-known Boltzmann solver CLASS \cite{Lesgourgues:2011rh}.  In this work, a simple power law parametrisation is used for $\mathcal{P}{\left( k \right)}$ given by
\begin{equation}
    \frac{k^3}{2 \pi^2} \mathcal{P}{\left( k \right)} = A \left( \frac{k}{\mathcal{K}} \right)^{n-1},
\end{equation}
with parameters $A \approx 2.1 \times 10^{-9}$, $n \approx 0.964$ consistent with the most recent reconstruction of the primordial power spectrum~\cite{Ade2016}, and the pivot scale $\mathcal{K}$ is set to the usual value of $0.05$/Mpc. We are specifically interested in the present-day value of \cref{eqn:PowerSpectrumDef} at $a = 1$, which can be obtained by solving \cref{eqn:PotentailEqn1} numerically and using the resulting present-day value of the potential $\phi$.

\begin{figure}[H]
    \centering
    \includegraphics[height=4.4cm]{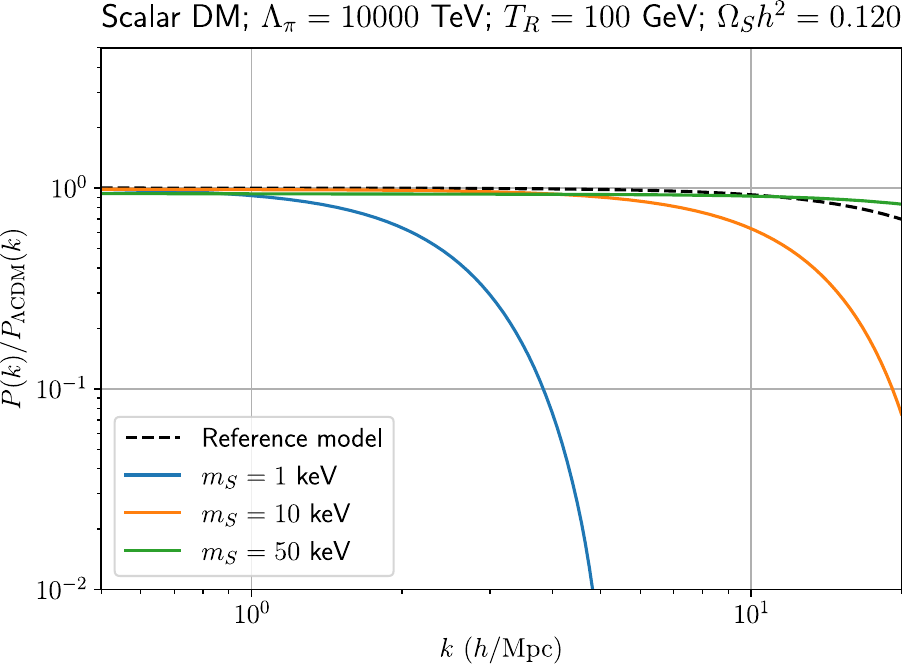}
    \includegraphics[height=4.4cm]{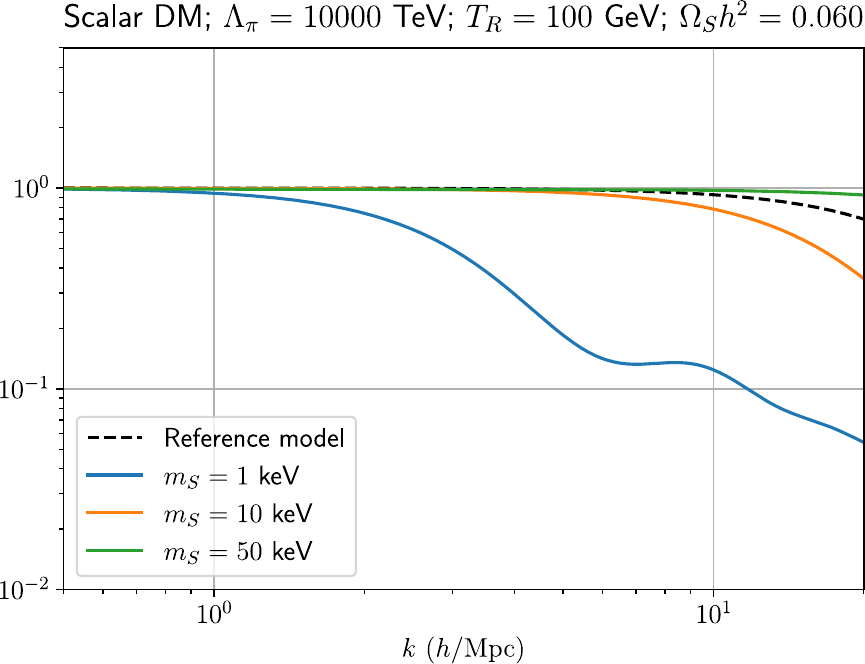}
    \includegraphics[height=4.4cm]{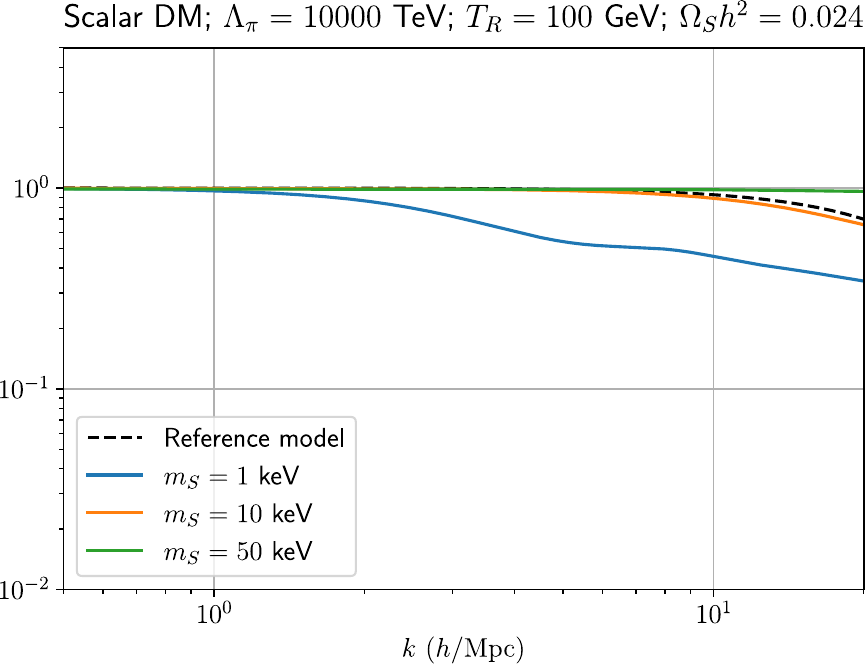}
    \caption{Plots of the transfer functions for the model parameters specified in the title card of the figures for various NCDM masses/fractions together with the reference model \cref{eqn:refModel}. $\Lambda_\pi$ is the cut-off scale of the theory, $T_R$ is the reheating temperature, $\Omega_S h^2$ is present day NCDM relic, and $m_S$ is the mass of the scalar dark matter candidate. Note that the qualitative difference between the resulting transfer functions for scalar and vector dark matter candidates is minimal, while the transfer function for the fermionic dark matter does not deviate significantly from the reference model due to the mass suppression of the collision functional. Thus, for brevity, we only present the transfer function for scalar NCDM species here.}
    \label{fig:PkPlotExpandTr}
\end{figure}

\begin{figure}
    \centering
    \includegraphics[height=5.1cm]{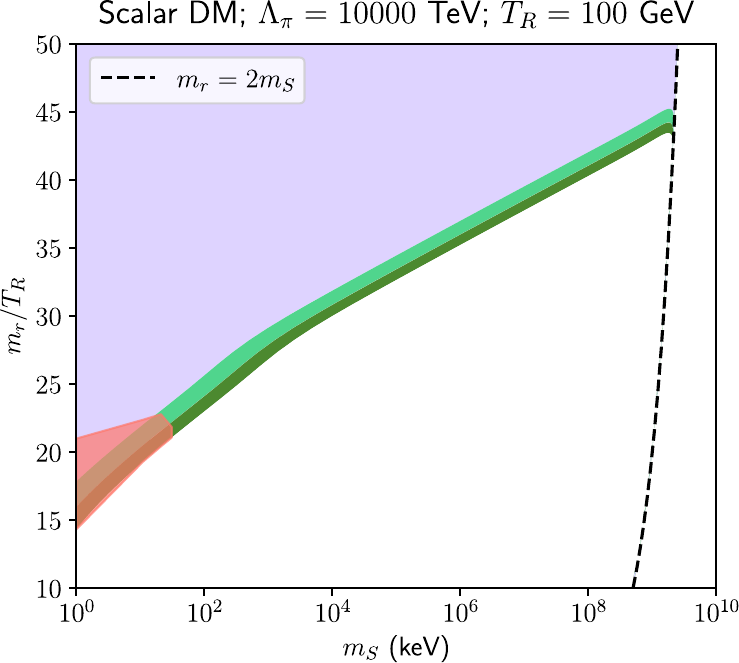}
    \includegraphics[height=5.1cm]{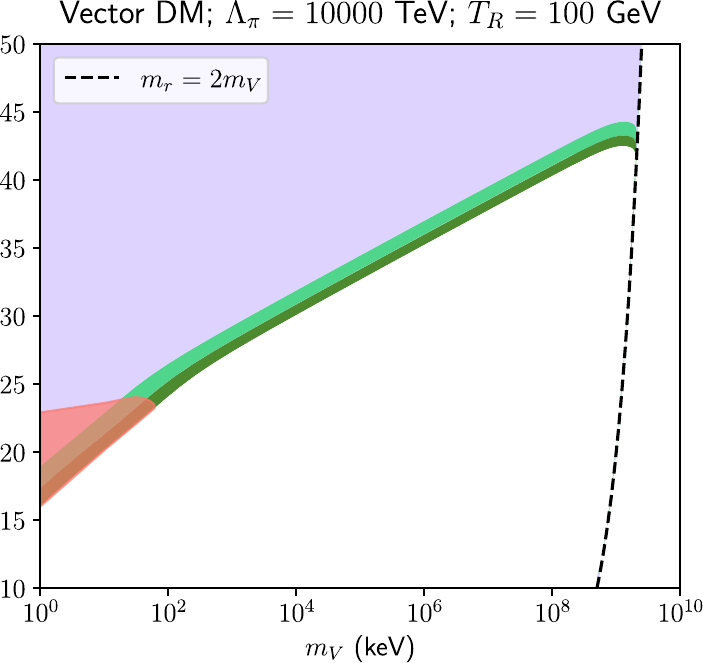}
    \includegraphics[height=5.1cm]{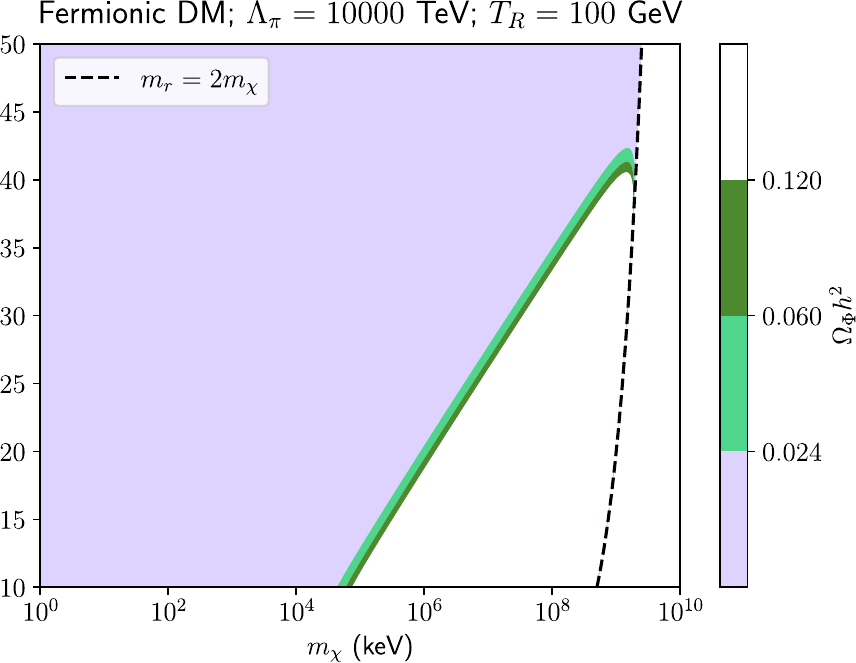} \\
    \caption{Parameter space of the model of dark matter species $\Phi = \{ S, V, \chi \}$ interacting with the standard model thermal bath through the radion portal. The plots are of the region of parameter space where $T_R \ll m_r$ such that the contribution from higher spin-$0$ KK modes can be neglected. $T_R$ is the reheating temperature, $m_\Phi$ is the mass of the dark matter species, and $m_r$ is the radion mass. The cut-off scale of the theory $\Lambda_\pi$ is set to $10000$ TeV. The red shaded region corresponds to the parameter space excluded by the $Ly$-$\alpha$ criteria $\delta A$ exceeding the reference value of $0.16$. \label{fig:RadionMediatorParameterSpaceScan}}
\end{figure}

To be precisely accurate, one needs to perform a full statistical analysis of the $Ly$-$\alpha$ data using the PSD that we computed to obtain absolute mass bounds. A simple way to assess the deviation of the obtained matter power spectrum from a thermal warm dark matter reference model through the area criterion, also known as the $\delta A$ criterion \
\cite{Murgia:2017lwo}\cite{Schneider:2016uqi}. We utilise the $1$D projection of the power spectrum defined as
\begin{equation}
    P_{1\text{D}}{\left( k \right)} = \int_k^{k_{\text{max}}} \frac{d k'}{2 \pi} k' P{\left( k' \right)}, \label{eqn:PowerSpectraProject}
\end{equation}
and the corresponding ratio integral defined as
\begin{equation}
    A = \int_{k_{\text{min}}}^{k_{\text{max}}} dk \frac{P_{1 \text{D}}{\left( k \right)}}{P_{1 \text{D}}^{\Lambda\text{CDM}}{\left( k \right)}},
\end{equation}
with $P_{1 \text{D}}^{\Lambda\text{CDM}}{\left( k \right)}$ obtained by applying \cref{eqn:PowerSpectraProject} to the power spectrum obtained from $\Lambda$CDM, in order to obtain the quantity
\begin{equation}
    \delta A = \frac{A_{\Lambda \text{CDM}} - A}{A_{\Lambda \text{CDM}}},
\end{equation}
with $A_{\Lambda \text{CDM}} = k_{\text{max}} - k_{\text{min}}$ by construction. For the purposes of this work, we chose to impose UV and IR cutoffs on the power spectrum consistent with the bounds of typical Lyman-$\alpha$ observations: $k_\text{min} = 0.5 h$/Mpc and $k_{\text{max}} = 20 h$/Mpc. As a reference model, we utilise a simple WDM model with $m_{\text{WDM}} = 3.5$ keV parametrised as
\begin{equation}
    P{\left( k \right)} = \left[ 1 + \left( \alpha k \right)^{2\nu} \right]^{- 10 / \nu} P_{\Lambda \text{CDM}}{\left( k \right)},\label{eqn:refModel}
\end{equation}
where $\nu = 1.12$ and $\alpha \approx 0.012h^{-1}$ Mpc~\cite{Viel2005,Viel2017}. The parametric function, also known as the \textit{transfer function}, is shown in \cref{fig:PkPlotExpandTr} alongside the numerically tabulated transfer function of our model. Upon application of the above prescription to the reference WDM model, we get the corresponding reference value of $\delta A_{\text{Ref}} \approx 0.16$, which serves as an estimate of the minimal power spectrum suppression for models that are excluded at the $95\%$ confidence level by the Lyman-$\alpha$ measurements. From \cref{fig:PkPlotExpandTr}, we see that there's a sharp drop-off in the power spectrum at small scales with the inclusion of the ``non-cold'' species, and this drop-off is more pronounced for lower NCDM masses and higher NCDM fractions of the relic. We also note that the $50$ \text{keV} green line lies above the reference model at small scales (higher $k$), implying that it is compatible with Lyman-$\alpha$ while lower DM masses (such as $10$ keV and $1$ keV) are clearly excluded. 
\cref{fig:RadionMediatorParameterSpaceScan} shows the parameter space of our model at $T_R = 100$ GeV and $\Lambda_\pi = 10000$ TeV in the $m_r \gg T_R$ regime. The red-shaded part indicates the region excluded by $Ly$-$\alpha$ constraints. There are a few important observations we could make:
\begin{itemize}
\item Due to the DM mass suppression in the collision functional, the relevant parameter space is unconstrained by $Ly$-$\alpha$ observations for the fermionic dark matter candidate.
\item The lower limit on the allowed mass is proportional to a relic fraction. 
Physically, this can be interpreted as scenarios with a higher fraction of non-cold dark matter requiring these non-cold species to be colder to be consistent with the $Ly$-$\alpha$ measurements.
\end{itemize}

\section{Conclusion}
\label{sec:conclusion}

In this work, we have computed the cosmological relic abundance of dark matter in a radion/dilaton mediated scenario produced via the freeze-in mechanism within a warped, stabilised Randall-Sundrum model. With the assumption that dark matter and the standard model particles are bound to the TeV brane, we show that keV-TeV scale dark matter particles (scalars, vectors, and fermions) can saturate the relic abundance. 

Relative to previous works on this subject, this paper computes the relevant scattering cross sections involving the radion mediator used to evaluate the final relic density in a fully consistent model of stabilized extra-dimensions with a massive scalar radion mediator. The calculation ensures that unitarity issues that have previous plagued calculations are avoided. Furthermore, instead of the usual number-density approach, we consistently calculate the relic density from the full phase-space distribution, thus reliably accounting for the non-thermal nature of the interaction and, therefore, the distribution function. Our results show that the commonly used criterion—namely, that the energy density of the produced dark matter is much smaller than its equilibrium value—does not justify neglecting the backward contribution to the collision term. In fact, we find that the backward process remains significant for cut-off scales of $\order{10000}$ TeV, even when the dark matter energy density is as low as $1\%$ of the equilibrium value. \looseness=-1

We also impose constraints on the model originating from Lyman-$\alpha$ data. Our analysis of the $Ly$-$\alpha$ constraints using the area criterion within the phenomenologically relevant limit $T_R \gg m_\Phi \gg m_\zeta$ 
excludes the models of the scalar and vector dark matter of $m_\Phi \lessapprox 100$ keV while leaving the fermionic dark matter unconstrained due to a power-law dependency of the collision functional for the fermionic dark matter on its mass. \looseness=-1

Finally, we compare our setup to the case in which dark matter is produced solely through gravitational interactions mediated by a massless spin-2 graviton (as would be the case in Einstein's general relativity in 4 spacetime dimensions). For the massless spin-2 mediator, the weakness of the gravitational coupling, relative to the reduced scale of compactified extra-dimensions, means that the relic density can be satisfied for extremely high reheating temperatures. For the phenomenologically relevant regime where $T_R \gg m_\Phi \gg m_\zeta$ with $\Phi$ denoting the dark matter species and $\zeta$ denoting various standard model species, the observed relic density within is saturated for the dark matter masses of $m_\Phi \simeq 10$ GeV $-$ $1000$ TeV for reheating temperatures of $T_R \simeq 10^{-4} M_{\text{Pl}} - 10^{-2} M_{\text{Pl}}$. Due to the relative weakness of the gravitational coupling, the only relevant constraints on this model are the ones arising from cosmological constraints on the reheating temperature for a given inflationary model, such as the ones arising from Planck data.

The radion portal set up therefore serves as a simple model of dark matter production via the freeze-in mechanism originating from higher dimensional gravitational interactions. Here we note that this work can be extended in a number of ways to generalize the set-up. First of all, as we have alluded to, if the reheating temperature is above the radion mass, the radion will be produced on shell from the inflationary bath. Since the same coupling determines the self-interactions in the gravitational sector and the interactions between the (dark) matter and the gravitational sector, we will require a consistent calculation of a coupled Boltzmann equation. Depending on the scale of the effective coupling, we can be in a regime of a two-component FIMP, or a Superwimp, or a combination of the two. Such scenarios may leave striking cosmological signatures, especially in $\rm Lyman-\alpha$ data. An even more complicated scenario can arise if the reheating temperature is higher than the number of KK gravitons which can be produced on-shell, in which case we have to consistently evaluate the contribution originating from these particles as well. Incidentally, these are reminiscent of freeze-in clockwork models \cite{Goudelis:2018xqi}, or continuum dark matter \cite{Csaki:2021gfm}.  Finally, we have not considered the dynamics of the inflaton and how it couples to dark matter via higher-dimensional gravitational interactions, including relaxing the assumption of instantaneous reheating. These issues require further work and will be part of forthcoming papers.

\section{Acknowledgement}

GS acknowledges the support he has received for his research through the provision of an Australian Government Research Training Program Scholarship and the support provided by the Adelaide University and the Australian Research Council through the Centre of Excellence for Dark Matter Particle Physics (CE200100008). GS is thankful to Prof. Anthony Williams for helpful and fruitful discussions. AM acknowledges the support received for her research through the provision of a Tuition Fee Scholarship (TFS) from Dr. Dipan Sengupta and the School of Physics, UNSW. AM and DS are thankful to Prof. Yvonne Y.Y. Wong for her suggestions and helpful discussions. AM, DS, and GS thank Prof. R. Sekhar Chivukula and Prof. Elizabeth Simmons for insightful conversations.

\appendix
\label{sec:feynrules}
\section{Feynman Rules and Matrix Elements}

\subsection{Graviton Mediator}
\label{sec:GraviFeyn}
\cref{eqn:gravactioninterterms} allows us to write down the following set of Feynman rules
\begin{align}
\vcenter{\hbox{\includegraphics[height=2.13cm]{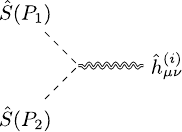}}} \hspace{0.07cm} & ~=~ \frac{i}{M_{\text{Pl}}} \left( P_1^\mu P_2^\nu + P_1^\nu P_2^\mu - \eta^{\mu \nu} \left( P_1 \cdot P_2 + m_S^2 \right) \right) , \label{eqn:GravitonScalarVertex}\\
\vcenter{\hbox{\includegraphics[height=2.13cm]{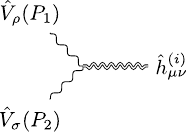}}} \hspace{0.07cm} & ~=~ - i \frac{1}{M_{\text{Pl}}} \left[ \left( m_V^2 + P_1 \cdot P_2 \right) C^{\mu \nu \rho \sigma} + \eta^{\mu \nu} P_1^\sigma P_2^\rho + \eta^{\sigma \rho} P_1^\nu P_2^\mu + \right. \nonumber \\ &\hspace{1cm} \left. + \eta^{\sigma \rho} P_1^\mu P_2^\nu  - \eta^{\mu \rho} P_1^\sigma P_2^\nu - \eta^{\mu \sigma} P_1^\nu P_2^\rho - \eta^{\nu \rho} P_1^\sigma P_2^\mu \right. \nonumber \\ &\hspace{1cm} - \left. \eta^{\nu \sigma} P_1^2 P_2^\rho \right], \label{eqn:GravitonVectorVertex} \\
\vcenter{\hbox{\includegraphics[height=2.13cm]{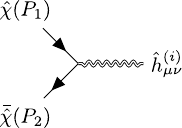}}} \hspace{0.07cm} & ~=~ - \frac{i}{4} \frac{1}{M_{\text{Pl}}} \left[ \gamma^\mu \left( P_1^\nu - P_2^\nu \right) + \gamma^\nu \left( P_1^\mu - P_1^\nu \right) - \right. \nonumber \\ &\hspace{1cm} \left. - 2 \eta^{\mu \nu} \left( \slashed{P}_1 - \slashed{P}_2 - 2 m_\chi \right) \right], \label{eqn:GravitonFermionVertex}
\end{align}
where all terms are defined identically to \cref{eqn:RadionScalarVertex,eqn:RadionVectorVertex,eqn:RadionFermionVertex}, and 
\begin{equation}
 C^{\mu \nu \alpha \beta} = \eta^{\mu \alpha} \eta^{\nu \beta} + \eta^{\mu \beta} \eta^{\nu \alpha} - \eta^{\mu \nu} \eta^{\alpha \beta}.
\end{equation}
The graviton field propagates as a massless spin-$2$ mode, with the corresponding propagator being
\begin{equation}
\vcenter{\hbox{\includegraphics[height=0.93cm]{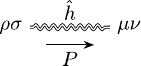}}} \hspace{0.07cm} ~=~ \frac{i}{2} \frac{1}{P^2} C^{\mu \nu \rho \sigma}. \label{eqn:GravitonPropogator}
\end{equation}

With the help of \cref{eqn:GravitonScalarVertex,eqn:GravitonVectorVertex,eqn:GravitonFermionVertex,eqn:GravitonPropogator} we can compute the corresponding $S$-matrix elements squared of the process depicted on \cref{fig:GravitonnPortalDiag}. For scalar dark matter candidate $\Phi = S$ we get:
\begin{align}
\abs{\mathcal{M}_{HH \rightarrow S S}}^2 &= \frac{1}{M_\text{Pl}^4 s^2} \left[ m_S^4 + m_H^4 + t \left( s + t \right)  - \left( m_S^2 + m_H^2 \right) \left( s + 2 t \right) \right]^2,  \label{eqn:SmatrixGHH} \\
\abs{\mathcal{M}_{ZZ \rightarrow S S}}^2 &= \frac{1}{M_\text{Pl}^4 s^2} \left[ 3 m_S^8 + 3 \left( m_Z^2 - t \right)^2 \left( s + t - m_Z^2 \right)^2 - 2 m_S^6 \left( s + 6 t \right) - \right. \nonumber \\
&\hspace{2cm}  2 m_S^2 \left( m_Z^4 \left( s + 6 t \right) + t \left( s + t \right) \left( s + 6 t \right) + m_Z^2 \left( s ^2 - 8 s t - 12 t^2 \right) \right) + \nonumber \\
&\hspace{4cm}  \left. m_S^4 \left( 6 m_Z^4 + s^2 + 2 m_Z^2 \left( s - 6 t \right) + 10 s t + 18 t^2 \right) \right],  \label{eqn:SmatrixGZZ} \\
\abs{\mathcal{M}_{W^+ W^- \rightarrow S S}}^2 &= \frac{1}{M_\text{Pl}^4 s^2} \left[ 3 m_S^8 + 3 \left( m_W^2 - t \right)^2 \left( s + t - m_W^2 \right)^2 - 2 m_S^6 \left( s + 6 t \right) - \right. \nonumber \\
&\hspace{2cm} 2 m_S^2 \left( m_W^4 \left( s + 6 t \right) + t \left( s + t \right) \left( s + 6 t \right) + m_W^2 \left( s ^2 - 8 s t - 12 t^2 \right) \right) + \nonumber \\
&\hspace{4cm} \left. m_S^4 \left( 6 m_W^4 + s^2 + 2 m_W^2 \left( s - 6 t \right) + 10 s t + 18 t^2 \right) \right],  \label{eqn:SmatrixGWW} \\
\abs{\mathcal{M}_{f \bar{f} \rightarrow SS}}^2 &= \frac{1}{2 M_\text{Pl}^4 \left( s - 4 m_f^2 \right) s^2} \left[ m_f^2 \left( 4 m_S^4 + \left( s + 2 t - 2 m_f^2 \right)^2 - 2 m_S^2 \left( s + 4 t \right)\right)^2 - \right. \nonumber \\ 
&\hspace{2cm} \left.  s \left( s + 2 \left( t -  m_S^2 - m_f^2 \right) \right)^2 \left( m_S^4 + m_f^4 - 2 m_f^2 t - 2 m_S^2 \left( m_f^2 + t \right) + t \left( s + t \right) \right)  \right],  \label{eqn:SmatrixGFF} \\
\abs{\mathcal{M}_{\gamma \gamma \rightarrow S S}}^2 &= \frac{1}{8} \abs{\mathcal{M}_{g g \rightarrow S S}}^2 = \frac{2}{M_\text{Pl}^4 s^2} \left( m_S^4 - 2 m_S^2 t + t \left( s + t \right) \right)^2, \label{eqn:SmatrixGPP}
\end{align}
where $s$ and $t$ are the Mandelstam variables. For the vector dark matter candidate, $\Phi = V$ we have
\begin{align}
    \abs{\mathcal{M}_{H H \rightarrow V V}}^2 &= \frac{1}{M_{\text{Pl}}^4 s^2} \left[ 3 \left( m_H^8 + m_V^8 \right) + 3 t^2 \left( s + t \right)^2 - 6 m_V^6 \left( s + 2 t \right) - 2 m_H^6 \left( s + 6 t \right) - \right. \nonumber \\ &\hspace{2cm}   2 m_H^2 t \left( s + t \right) \left( s + 6 t \right) + m_H^4 \left( s^2 + 10 s t + 18 t^2 \right) + \nonumber \\ &\hspace{4cm} 2 m_V^2 \left( m_H^2 - s - t \right) \left( m_S^2 \left( s - 6 t \right) + 3 t \left( s + 2 t \right) \right) + \nonumber \\ &\hspace{6cm} \left. m_V^4 \left( 6 m_H^6 - 2 m_H^2 \left( s + 6 t \right) + 3 \left( s^2  +6 s t + 6 t^2 \right) \right) \right], \label{eqn:SmatrixGSSVector} \\
    \abs{\mathcal{M}_{ZZ \rightarrow VV}}^2 &= \frac{1}{M_{\text{Pl}}^4 s^2} \left[2 s^4 + 9 t^4 + 9 \left( m_V^4 + m_Z^4 \right)^2 - 6 \left( m_V^2 + m_Z^2 \right) \left( m_V^4 - 4 m_V^2 m_Z^2 + m_Z^4 \right) s +\right. \nonumber \\ &\hspace{2cm} + 17 s^2 t^2 +  s^2 \left( 11 m_V^4 + 18 m_V^2 m_Z^2 + 11 m_Z^4 \right) + 18 t^3 \left( s - 2 \left( m_V^2 + m_Z^2 \right) \right) - \nonumber \\ &\hspace{4cm} 2 \left( 2 \left( m_V^2 + m_Z^2 \right) - s \right) \left( 9 \left( m_V^4 + m_Z^4 \right) - 3 \left( m_V^2 + m_Z^2 \right) s + 4 s^2 \right) t + \nonumber \\ &\hspace{5cm}\left.  \left( 18 \left( 3 m_V^4 + 4 m_V^2 m_Z^2 + 3 m_Z^4 \right) - 42 s \left( m_V^2 + m_Z^2 \right) s \right) t^2 \right], \label{eqn:SmatrixGZZVector} \\
    \abs{\mathcal{M}_{W^+ W^- \rightarrow VV}}^2 &= \frac{1}{M_{\text{Pl}}^4 s^2} \left[2 s^4 + 9 t^4 + 9 \left( m_V^4 + m_W^4 \right)^2 - 6 \left( m_V^2 + m_W^2 \right) \left( m_V^4 - 4 m_V^2 m_W^2 + m_W^4 \right) s + \right. \nonumber \\ &\hspace{2cm} + 17 s^2 t^2 +  s^2 \left( 11 m_V^4 + 18 m_V^2 m_W^2 + 11 m_W^4 \right) + 18 t^3 \left( s - 2 \left( m_V^2 + m_W^2 \right) \right) - \nonumber \\ &\hspace{4cm} 2 \left( 2 \left( m_V^2 + m_W^2 \right) - s \right) \left( 9 \left( m_V^4 + m_W^4 \right) - 3 \left( m_V^2 + m_W^2 \right) s + 4 s^2 \right) t + \nonumber \\ &\hspace{5cm}\left.  \left( 18 \left( 3 m_V^4 + 4 m_V^2 m_W^2 + 3 m_W^4 \right) - 42 s \left( m_V^2 + m_W^2 \right) s \right) t^2 \right], \label{eqn:SmatrixGWWVector} \\
    \abs{\mathcal{M}_{f \bar{f} \rightarrow VV} }^2  &=  \frac{1}{2 M_{\text{Pl}}^4 s^2} \left[  12 m_V^6 \left( s+4t \right) - 12 m_V^8 - m_V^4 \left( 24 m_f^4 + 4 m_f^2 \left( s - 12t \right) + 5s^2 + 48st + 72 t^2 \right) - \right. \nonumber \\ &\hspace{2cm} 2 m_V^2 \left(2 m_f^4 \left( 5s-12t \right) + m_f^2 \left( s^2 + 20st + 48 t^2 \right) - 2 s^3 - 11 s^2 t - 30 s t^2 - 24 t^3 \right) - \nonumber \\ &\hspace{3cm}  - t (s+t) \left(5 s^2+12 s t+12 t^2\right) + 8 m_f^6 \left( s+6 t \right) - m_f^4 \left(9 s^2 + 40 s t + 72 t^2\right) - \nonumber \\ &\hspace{5cm} \left. m_f^2 \left( s+2 t \right) \left(s^2+16 s t+24 t^2\right) -   12 m_f^8 \right], \label{eqn:SmatrixGffVector} \\
    \abs{\mathcal{M}_{\gamma \gamma \rightarrow VV}}^2 &= \frac{1}{8} \abs{\mathcal{M}_{g g \rightarrow V V}}^2 = \frac{2}{M_{\text{Pl}}^2 s^2} \left[ m_V^2 + s^2 - 2 m_V^2 t + t \left( s + t \right) \right] \left[ s^2 + 3 \left( m_V^4 - 2 m_V^2 t + t \left( s + t \right) \right) \right]. \label{eqn:SmatrixGgammagammaVector}
\end{align}
Lastly, for the fermion dark matter candidate $\Phi = \chi$, we have 
\begin{align}
    \abs{\mathcal{M}_{HH \rightarrow \chi \bar{\chi}}}^2 &= \frac{1}{2 M_{\text{Pl}}^4 s^2} \left[  8 m_\chi^6 \left( s+2 t \right) - 4 m_\chi^8 - m_\chi^4 \left(8 m_H^4 - 4 m_H^2 \left( s + 4 t \right) + 5s^2 + 24 t \left( s +  t \right) \right) + \right. \nonumber \\ &\hspace{2cm} m_\chi^2 \left(4 m_H^4 \left( s + 4 t \right) - 2 m_H^2 \left( s^2 + 12 s t + 16 t^2 \right) + \left( s+ 2 t \right) \left(s^2+8 s t+8 t^2\right)\right) - \nonumber \\ &\hspace{4cm} \left. \left(s +2 t -2 m_H^2 \right)^2 \left(m_H^4-2 m_H^2 t+ \left( s + t \right)t \right) \right], \label{eqn:SmatrixGHHFermion} \\
    \abs{\mathcal{M}_{Z Z \rightarrow \chi \bar{\chi}}}^2 &= \frac{1}{2 M_{\text{Pl}}^4 s^2} \left[  8 m_\chi^6 \left( s + 6 t \right) - 12 \left( m_\chi^8 + m_Z^8 \right) - m_\chi^4 \left(24 m_Z^4+4 m_Z^2 \left(5s-12t \right)+9 s^2+40 s t+72 t^2\right) - \right. \nonumber \\ &\hspace{2cm} m_\chi^2 \left(4 m_Z^4 \left( s - 12 t \right)+2 m_Z^2 \left(s^2+20 s t+48 t^2\right) - \left( s + 2 t \right) \left(s^2+16 s t+24 t^2\right)\right)  - \nonumber \\ &\hspace{4cm} m_Z^4 \left(5 s^2+48 s t+72 t^2\right) + 2 m_Z^2 \left(2 s^3+11 s^2 t+30 s t^2+24 t^3\right) - \nonumber \\ &\hspace{5cm} \left. t \left(s+t \right) \left(5 s^2+12 s t+12 t^2\right) +  12 m_Z^6 \left( s + 4 t \right) \right], \label{eqn:SmatrixGZZFermion} \\
    \abs{\mathcal{M}_{W^+ W^- \rightarrow \chi \bar{\chi}}}^2 &= \frac{1}{2 M_{\text{Pl}}^4 s^2} \left[  8 m_\chi^6 \left( s + 6 t \right) - 12 \left( m_\chi^8 + m_W^8 \right) - m_\chi^4 \left(24 m_W^4+4 m_W^2 \left(5s-12t \right)+9 s^2+40 s t+72 t^2\right) - \right. \nonumber \\ &\hspace{2cm} m_\chi^2 \left(4 m_W^4 \left( s - 12 t \right)+2 m_W^2 \left(s^2+20 s t+48 t^2\right) - \left( s + 2 t \right) \left(s^2+16 s t+24 t^2\right)\right)  - \nonumber \\ &\hspace{4cm} m_W^4 \left(5 s^2+48 s t+72 t^2\right) + 2 m_W^2 \left(2 s^3+11 s^2 t+30 s t^2+24 t^3\right) - \nonumber \\ &\hspace{5cm} \left. t \left(s+t \right) \left(5 s^2+12 s t+12 t^2\right) +  12 m_W^6 \left( s + 4 t \right) \right], \label{eqn:SmatrixGWWFermion} \\
    \abs{\mathcal{M}_{f \bar{f} \rightarrow \chi \bar{\chi}}}^2 &= \frac{1}{8 M_{\text{Pl}}^4 s^2} \left[32 \left(m_\chi^4+m_f^4\right)^2 - 32 s \left(m_\chi^6+m_f^6\right)-4 s^3 \left(m_\chi^2+m_f^2\right)-64 t^3 \left(2 \left(m_\chi^2+m_f^2\right)-s\right) + \right. \nonumber \\ &\hspace{2cm} 2 t^2 \left(-80 s \left(m_\chi^2+m_f^2\right)+32 \left(3 m_\chi^4+4 m_\chi^2 m_f^2+3 m_f^4\right)+21 s^2\right)- \nonumber \\ &\hspace{4cm} 2 t \left(2 \left(m_\chi^2+m_f^2\right)-s\right) \left(32 \left(m_\chi^4+m_f^4\right)-16 s \left(m_\chi^2+m_f^2\right)+5 s^2\right)+ \nonumber \\ &\hspace{4cm} \left. 2 s^2 \left(5 m_\chi^4+2 m_\chi^2 m_f^2+5 m_f^4\right)+s^4+32 t^4 \right], \label{eqn:SmatrixGffFermion} \\
    \abs{\mathcal{M}_{\gamma \gamma \rightarrow \chi \bar{\chi}}}^2 &= \frac{1}{8} \abs{\mathcal{M}_{g g \rightarrow \chi \bar{\chi}}}^2 = \frac{2}{M_{\text{Pl}}^2 s^2} \left[ 2 m_\chi^2 t - m_\chi^4 - t \left( s + t \right) \right] \left[ s^2 + 2 \left( m_\chi^4 - 2 m_\chi^2 t + t \left( s + t \right) \right) \right]. \label{eqn:SmatrixGgammagammaFermion}
\end{align}

\subsection{Radion Mediator}
\cref{eqn:actioninterterms} leads to the following set of Feynman rules 
\begin{align}
\vcenter{\hbox{\includegraphics[height=2.13cm]{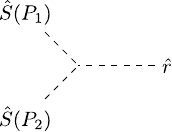}}} \hspace{0.07cm} & ~=~ i\sqrt{\frac{2}{3}} \frac{1}{\Lambda_\pi} \left( 2 m_S^2 + P_1 \cdot P_2 \right), \label{eqn:RadionScalarVertex}\\
\vcenter{\hbox{\includegraphics[height=2.13cm]{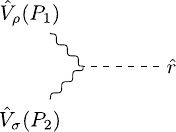}}} \hspace{0.07cm} & ~=~ - i\sqrt{\frac{2}{3}} \frac{1}{\Lambda_\pi} m_V^2 \eta^{\rho \sigma}, \label{eqn:RadionVectorVertex} \\
\vcenter{\hbox{\includegraphics[height=2.13cm]{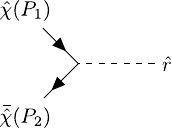}}} \hspace{0.07cm} & ~=~ - i\sqrt{\frac{2}{3}} \frac{1}{\Lambda_\pi} \left[ \frac{3}{4} \left( \slashed{P}_1 + \slashed{P}_2 \right) - 2 m_\chi \right], \label{eqn:RadionFermionVertex}
\end{align}
where $P_1$ and $P_2$ are the incoming four-momenta, $\slashed{K} = \gamma^\mu K_\mu$ where $\gamma^\mu$ is the Dirac $\gamma$-matrix, $\eta_{\mu \nu}$ is the usual Minkowski metric, labels $S$, $V$, and $\chi$ are placeholders for the scalar, vector, and fermion species respectively with the factors $m_S$, $m_V$, and $m_\chi$ corresponding to the scalar, vector, and fermion masses. Note that Feynman rules \cref{eqn:RadionScalarVertex,eqn:RadionVectorVertex,eqn:RadionFermionVertex} apply to both dark and standard model sectors. The radion field propagates as a massive spin-$0$ mode, with the corresponding propagator being
\begin{equation}
\vcenter{\hbox{\includegraphics[height=0.93cm]{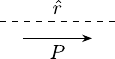}}} \hspace{0.07cm} ~=~ \frac{i}{P^2 - m_r^2 + i m_r \Gamma_r}, \label{eqn:RadionProp}
\end{equation}
where $m_r$ is the radion mass and $\Gamma_r$ is the total decay width of the radion. For the cut-off scales $\Lambda_\pi$ considered in this work, it is safe to assume that $\Gamma_r / m_r \ll 1$, allowing us to express the square of the radion propagator in terms of the Dirac $\delta$-function through the narrow width approximation
\begin{equation}
     \frac{1}{\left( s - m_r \right)^2 + m_r^2 \Gamma_r^2} \approx \frac{\pi}{m_r \Gamma_r} \delta{\left( s- m_r^2 \right)}. \label{eqn:RadionNWA}
\end{equation}
With the help of \cref{eqn:RadionScalarVertex,eqn:RadionVectorVertex,eqn:RadionFermionVertex,eqn:RadionProp,eqn:RadionNWA} we can write down the squares of $S$-matrix elements corresponding to the process depicted on \cref{fig:RadionPortalDiag}:
\begin{align}
    \abs{\mathcal{M}_{H H \rightarrow \Phi \Phi}}^2 &= \frac{16 \pi^2}{3 \Lambda_\pi^2} \left[ \sqrt{ 1 - \frac{4 m_H^2}{m_r^2} } \right]^{-1} \upsilon_\Phi{\left( m_r, m_\Phi \right)} \beta{\left( r \rightarrow H  H \right)} \delta{\left( s - m_r^2 \right)}, \label{eqn:SmatrixHHChiChi} \\
    \abs{\mathcal{M}_{Z Z \rightarrow \Phi \Phi}}^2 &= \frac{16 \pi^2}{3 \Lambda_\pi^2} \left[ \sqrt{ 1 - \frac{4 m_Z^2}{m_r^2} } \right]^{-1}  \upsilon_\Phi{\left( m_r, m_\Phi \right)} \beta{\left( r \rightarrow Z  Z \right)} \delta{\left( s - m_r^2 \right)}, \label{eqn:SmatrixZZChiChi} \\ 
    \abs{\mathcal{M}_{W^+ W^- \rightarrow \Phi \Phi}}^2 &= \frac{8 \pi^2}{3 \Lambda_\pi^2} \left[ \sqrt{ 1 - \frac{4 m_W^2}{m_r^2} } \right]^{-1}  \upsilon_\Phi{\left( m_r, m_\Phi \right)} \beta{\left( r \rightarrow W^+  W^- \right)} \delta{\left( s - m_r^2 \right)}, \label{eqn:SmatrixWWChiChi} \\
    \abs{\mathcal{M}_{f  \bar{f} \rightarrow \Phi \Phi}}^2 &= \frac{8 \pi^2}{3 \Lambda_\pi^2} \left[ \sqrt{ 1 - \frac{4 m_f^2}{m_r^2} } \right]^{-1}  \upsilon_\Phi{\left( m_r, m_\Phi \right)} \beta{\left( r \rightarrow f  \bar{f} \right)} \delta{\left( s - m_r^2 \right)}, \label{eqn:SmatrixFFChiChi}
\end{align}
where $\beta{\left( r \rightarrow \zeta \zeta \right)}$ is the branching ratio of the radion into the species $\zeta$, $m_f$ is the placeholder for the mass of various standard model fermions, while $m_H$, $m_Z$, $m_W$ are the Higgs, $Z$-boson, and $W$-boson masses respectively, $m_\Phi$ is the mass of the dark matter candidate, $s$ is the $s$ Mandelstam variable, and $ \upsilon_\Phi{\left( m_r, m_\Phi \right)}$ are the interaction dependant factors defined as 
\begin{align}
    \upsilon_S{\left( m_r, m_S \right)} &= \left( m_r^2  + 2 m_S^2 \right)^2, \label{eqn:upsilonS} \\
    \upsilon_V{\left( m_r, m_V \right)} &= m_r^4 - 4 m_r^2 m_V^2 + 12 m_V^4, \label{eqn:upsilonV} \\
    \upsilon_\chi{\left( m_r, m_\chi \right)} &= 2 m_\chi^2 \left( m_r^2 - 4 m_\chi^2 \right),\label{eqn:upsilonChi}
\end{align}
which differ or scalar $\left( S \right)$, vector $\left( V \right)$, and fermion $\left( \chi \right)$ dark matter candidates. Note that to obtain \cref{eqn:SmatrixHHChiChi,eqn:SmatrixZZChiChi,eqn:SmatrixWWChiChi,eqn:SmatrixFFChiChi} we have implicitly assumed that the radion is heavier than any of the standard model or dark matter species $\left( m_r > 2 m_\zeta \text{ and } m_r > 2 m_\Phi \right)$.

\section{Boltzmann Equation and Collision Functional}
\label{sec:colfunc}
We start by noting that the distribution function of species $f_\Phi$ obeys a well-known transport equation
\begin{equation}
    \frac{d f_\Phi}{dt} = C{\left[ f_\Phi \right]}, \label{eqn:FTransportEqn}
\end{equation}
where $t$ is the cosmic time. For the purposes of this work, we assume that the collision functional appearing on the right-hand side of \cref{eqn:FTransportEqn} is given entirely in terms of the collision functional described in \cref{sec:colfunc} as the only relevant interactions are between the standard model thermal bath and the dark matter thermal bath. To solve \cref{eqn:FTransportEqn}, we need to make certain assumptions about the function $f_\Phi$. We start by invoking the cosmological principle, which states that the distribution function $f_\Phi$ is homogeneous; hence, we can assume it depends only on the magnitude of the three-momentum vector of species $\Phi$, $p$, and cosmic time $t$. In other words, $f_\Phi = f_\Phi{\left( p, t \right)}$. Furthermore, as we are considering an expanding universe, we are working with the Friedmann–Lemaître–Robertson–Walker geometry where 
\begin{equation}
    \frac{dp}{dt} = -H{\left( t \right)} p, \label{eqn:FLRWDefinition}
\end{equation}
with $H{\left( t \right)}$ being the Hubble constant. With the application of the chain rule and \cref{eqn:FLRWDefinition}, we can write \cref{eqn:FTransportEqn} in a more familiar form
\begin{equation}
    \frac{\partial f_\Phi}{\partial t} - H{\left( t \right)} p \frac{\partial f_\Phi}{\partial p} = C{\left[ f_\Phi \right]}. \label{eqn:BoltzmanEqn}
\end{equation}
As it is more convenient to work with the temperature of the standard model bath $T$ rather than cosmic time $t$, we can use the fact that we are concerned with the production of species $\Phi$ in the early radiation-dominated universe to link the cosmic time $t$ to reduced temperature $x = m_\Phi / T$ as 
\begin{equation}
    \frac{dt}{dx} = \frac{\sqrt{90} M_{\text{Pl}}}{\pi \sqrt{g^*{\left( x \right)}} m_\Phi^2} x \text{ and } H{\left( t \right)} = \frac{\pi \sqrt{g^*{\left( x \right)}} m_\Phi^2}{\sqrt{90} M_{\text{Pl}} x^2}, \label{eqn:timetemperaturelink}
\end{equation}
where $g^*{\left( x \right)}$ is the number of open degrees of freedom of the standard model bath, $M_{\text{Pl}}$ is the Planck mass, and $m_\Phi$ is the mass of dark matter candidates $\Phi$. Lastly, due to the spherical symmetry of the problem, it is more numerically convenient to work with the distribution function $g_\Phi = p^2 f_\Phi$, which allows us to rewrite \cref{eqn:BoltzmanEqn} as 
\begin{equation}
    \frac{\partial g_\Phi}{\partial \mu} - \frac{\partial g_\Phi}{\partial \nu} = \frac{\sqrt{90}M_{\text{Pl}}  e^{2\left( \mu + \nu \right)}}{\pi \sqrt{g^*{\left( \mu \right)}}} \mathcal{C}{\left[ g_\Phi \right]} - 2 g_\Phi, \label{eqn:TransportEqnFull}
\end{equation}
where we have introduced new variables $\mu = \text{ln}{\left( x \right)}$ and $\nu = \text{ln}{\left( p / m_\Phi \right)}$. The function $g_\Phi{\left( \mu, \nu \right)}$ obeys Dirichlet boundary conditions $g_\Phi{\left( \mu, 0 \right)} = g_\Phi{\left( \mu, \infty \right)} = 0$ for all $\mu$ and the initial condition $g_\Phi{\left( \mu_0, \nu \right)} = 0$ for all $\nu$ for some initial value of $\mu_0 = \text{ln}{\left( m_\Phi / T_R \right)}$ where $T_R$ is the reheating temperature. 

The distribution function $f_\Phi{\left(p, T \right)}$ obtained by solving \cref{eqn:TransportEqnFull} numerically can be related to the relic abundance of the dark matter species at a given temperature of the standard model bath $T$ through the definition of the energy density
\begin{equation}
    \Omega_\Phi{\left( T \right)} = \frac{g^*_\Phi}{3 H{\left( T \right)}^2 M_\text{Pl}^2} \int \frac{d^3 p}{\left( 2 \pi \right)^3} \sqrt{p^2 + m_\Phi^2}  f_\Phi{\left( p, T \right)}, \label{eqn:OmegaDefinition}
\end{equation}
where $g^*_\Phi$ is the number of degrees of freedom carried by the DM species $\Phi$. \cref{eqn:OmegaDefinition} naively implies that in order to obtain the relic density of species $\Phi$ today, we would have to numerically evolve \cref{eqn:TransportEqnFull} from some large reheating temperature $T_R$ all the way down to the CMB temperature. This is, however, unnecessary, as we only need to evolve \cref{eqn:TransportEqnFull} up to a point where the production of the species $\Phi$ stops, as any evolution beyond this point is due to Hubble expansion and can be expressed as a simple redshift $f_\Phi{\left(p, T \right)} = f_\Phi{\left( a{\left( T \right)} p \right)}$ for all $T$ smaller than some $T^\star$ at which the production of species $\Phi$ stops. For the processes considered here, a good approximation for $T^\star$ is $T^\star = T_R / 4$. 

The general expression for the collision functional of the ``two to two'' annihilation process $1 + 2 \rightarrow 3 + \Phi$ is given by the well-known expression
\begin{multline}
    C{\left[ f_\Phi \right]} = \frac{1}{2 g^*_\Phi E_\Phi n!} \int \prod_{i=1}^3 \left( \frac{d^3 p_i}{2 E_i \left( 2 \pi \right)^3}\right)  \left( 2 \pi 
    \right)^4 \delta^{\left( 4 \right)}{\left( P_1 + P_2 - P_3 - P_\Phi \right)} \times \\  \left( \abs{\mathcal{M}_{1 + 2 \rightarrow 3 + \Phi}}^2 f_1 f_2 \left( 1 \pm f_3 \right) 
    \left( 1 \pm f_\Phi \right) -   \abs{\mathcal{M}_{3 + \Phi \rightarrow 1 + 2}}^2 f_3 f_\Phi \left( 1 \pm f_1 \right) \left( 1 \pm f_2 \right) \right), \label{eqn:CollFuncGeneral}
\end{multline}
where $E_i$, $P_i$, $f_i$ are the energy, four-momentum, and the distribution function of the species $i$, $\delta^{\left( 4 \right)}{\left( P_1 + P_2 - P_3 - P_\Phi \right)}$ is the four-momentum conserving $\delta$-function, $\mathcal{M}_{1 + 2 \rightarrow 3 + \Phi}$ and $\mathcal{M}_{3 + \Phi \rightarrow 1 + 2}$ are the $S$-matrix elements of the forward and backward process respectively, $\left( 1 \pm f_i \right)$ is the Bose enhancement or Pauli blocking factor arising due to the spin statistics of the species $i$, $g^*_\Phi$ is the number of degrees of freedom carried by species $\Phi$, and $n!$ is the combinatoric factor that ensures we do not double count the initial momentum configurations for the case of $n$ indistinguishable species in the initial state. 

To simplify \cref{eqn:CollFuncGeneral}, we use the fact that, by crossing symmetry, all of the matrix elements considered in this work are identical to those of the backwards process. Furthermore, we use the fact that the standard model bath stays in thermal equilibrium. Lastly, ignoring the spin statistics allows us to disregard the Bose enhancement and Pauli blocking factors and to set the distribution of the standard model species to the Maxwell-Boltzmann distribution. Following the approach outlined in Ref.~\cite{Du2022}, we are left with a rather simple expression for the collision functional
\begin{multline}
    C{\left[ f_\Phi \right]} = \frac{1}{512  \pi^2 g^*_\Phi E p n!} \int_{4 m_\Phi^2}^\infty \frac{ds}{p^*{\left( s \right)} \sqrt{s}} \int_{t^{\text{min}}{\left( s \right)}}^{t^{\text{max}}{\left( s \right)}} dt \abs{\mathcal{M}_{\zeta \zeta \rightarrow \Phi \Phi}}^2 \times \\ \left[ T e^{-E/ T} \left( e^{-E^{\text{min}}{\left( s \right)} / T} - e^{-E^{\text{max}}{\left( s \right)} / T}  \right) - f_\Phi \left( F{\left( E^{\text{max}}{\left( s \right)}\right)} - F{\left( E^{\text{min}}{\left( s \right)}\right)} \right) \right], \label{eqn:collisionfuncdef} 
\end{multline}
where $T$ is the temperature of a thermal bath of species $\zeta$, $n$ is the number of identical species $\zeta$, $E$ is the energy of species $\Phi$ given by the usual relativistic dispersion relation $E = \sqrt{p^2 + m_\Phi^2}$ with $p$ denoting the momentum of species $\Phi$ in the cosmic frame, $p^*{\left( s \right)}$ is the momentum of species $\Phi$ in centre of momentum frame given by
\begin{equation}
    p^*{\left( s \right)} = \frac{1}{2} \sqrt{s - 4 m_\Phi^2},
\end{equation}
$E^{\text{min}/\text{max}}{\left( s \right)}$ are the lower and upper bounds on the energy of the species $\Phi$ when the centre of momentum energy $s$ and the cosmic frame momentum $p$ are fixed 
\begin{equation}
    E^{\text{min}/\text{max}}{\left( s \right)} = \sqrt{ m_\Phi^2 + \frac{1}{4m_\Phi^4} \left( \sqrt{ s \left( m_\Phi^2 + p^2 \right) \left( s - 4 m_\Phi^2 \right) } \pm \left( 2 m_\Phi^2 p - p s \right) \right)^2 },
\end{equation}
and $t^{\text{min}/\text{max}}$ are the lower and upper bounds on the Mandelstam variable $t$ for the fixed centre of momentum energy $s$ given by
\begin{equation}
    t^{\text{min}/\text{max}}{\left( s \right)} = m_\Phi^2 + m_\zeta^2 - \frac{1}{2} s \mp \frac{1}{2} \sqrt{ \left( s - 4 m_\Phi^2 \right) \left( s - 4 m_\zeta^2 \right)}. \label{eqn:tminmaxdef}
\end{equation}
Lastly, we have a function $F{\left( E \right)}$ denoting the integral over the distribution function $f_\Phi$ 
\begin{equation}
    F{\left( E \right)} = \int_0^{\sqrt{E^2 - m_\Phi^2}}  \frac{dp p}{\sqrt{p^2 + m_\Phi^2}} f_\Phi.
\end{equation}

\section{Extracting Lyman-\texorpdfstring{$\alpha$}{alpha} constraints}
\label{sec:lya}
We work in the longitudinal/Newtonian gauge, where the geometry is parametrised as
\begin{equation}
    ds^2 = - \left( 1 + 2 \psi \right) dt^2 + a{\left( t \right)}^2 \left( 1 + 2 \phi \right) dx^i dx_i, \label{eqn:EinsteinEqnGeom}
\end{equation}
where $a{\left(t \right)}$ is the scale factor and $\psi/\phi$ are the perturbations of the metric. The power spectrum is directly proportional to the amplitude of metric perturbations $\phi$. Hence, to compute today's matter power spectrum that can be constrained from $Ly$-$\alpha$ measurements, we need to evolve the geometry \cref{eqn:EinsteinEqnGeom} from some initial condition $a{\left( t_0 \right)} = a_0 \ll 1$ up to today $a{\left( t_\text{today} \right)} = 1$ according to Einstein's equation.

To simplify the computation, the main approximation we will rely upon is that the evolution of the distribution function of species $\Phi$ is given purely in terms of the redshift, hence $f_\Phi{\left(p, T \right)} = f_\Phi{\left( a p \right)}$. In other words, we assume that the initial condition $a_0$ is at a much later cosmic time than the point at which the production of the species $\Phi$ stopped. Then we can parametrise the distribution function of species $\Phi$ at any point in space-time as 
\begin{equation}
    f_\Phi{\left( \vec{x}, \vec{p}, a \right)} = f_\Phi{\left( q \right)} \left( 1 + \omega{\left( \vec{x}, \vec{p}, a \right)} \right),
\end{equation}
where $q = a p$ is today's momentum, $\vec{p}$ is the three-momentum of species $\Phi$ with the magnitude $p$, $\vec{x}$ is the position in the cosmic frame, and $a$ is the scale factor that is used to parametrise time. The perturbation $\omega{\left( \vec{x}, \vec{p}, a \right)}$ is expanded in the Legendre series
\begin{equation}
    \omega{\left( \vec{x}, \vec{p}, a \right)} = \sum_{l=0}^\infty \left( - i \right)^l \left( 2 l + 1 \right) \omega_l{\left( q, a \right)} P_l{\left( \hat{p} \cdot \hat{x} \right)},
\end{equation}
where $\hat{p}$ and $\hat{x}$ are the unit vectors in the direction of $\vec{p}$ and $\vec{x}$ respectively, $P_l$ is the Legendre polynomial, and $\omega_l$ is the corresponding expansion coefficient. Hence, the quantities required to build the stress-energy tensor of the species $\Phi$ for any value of scale factor $a$ can be expressed in terms of the expansion coefficients $\omega_l$ as
\begin{align}
    \bar{\rho}_\Phi{\left( a \right)} &= \frac{g^*_\Phi}{2 \pi^2 a^4} \int dq q^2 \sqrt{q^2 + m_\Phi^2 a^2} f_\Phi{\left( q \right)}, \\
    \bar{P}_\Phi{\left( a \right)} &= \frac{g^*_\Phi}{6 \pi^2 a^4} \int \frac{dq q^4}{\sqrt{q^2 + m_\Phi^2 a^2}} f_\Phi{\left( q \right)} , \\
    \delta_\Phi{\left( a \right)} &= \frac{\int dq q^2 \sqrt{q^2 + m_\Phi^2 a^2} f_\Phi{\left( q \right)} \omega_0{\left( q, a \right)} }{\int dq q^2 \sqrt{q^2 + m_\Phi^2 a^2} f_\Phi{\left( q \right)}}, \\
    \left( \bar{\rho}_\Phi{\left( a \right)} + \bar{P}_\Phi{\left( a \right)} \right) \sigma_\Phi{\left( a \right)} &= \frac{g^*_\Phi}{3 \pi^2 a^4} \int  \frac{dq q^4}{\sqrt{q^2 + m_\Phi^2 a^2}} f_\Phi{\left( q \right)} \omega_2{\left( q, a \right)}, 
\end{align}
where $g^*_\Phi$ is the number of degrees of freedom carried by species $\Phi$, $\bar{\rho}_\Phi{\left( a \right)}$ is the energy density, $\bar{P}_\Phi{\left( a \right)}$ is the pressure, $\delta_\Phi{\left( a \right)}$ is a relative perturbation of the energy density, and $\sigma_\Phi{\left( a \right)}$ is the relative perturbative of the sheer stress. Furthermore, we express the Hubble constant according to the Friedmann equation as 
\begin{equation}
H{\left( a \right)} = H_0 \sqrt{ \Omega_{\Lambda0} + \left( \Omega_{\gamma0} + \Omega_{\nu 0} \right) / a^4 + \left( \Omega_{b0} + \Omega_{\text{CDM}0} \right)/a^3 + \Omega_{k0} / a^2 + \Omega_{\Phi0} \bar{\rho}_\Phi{\left( a \right)} / \bar{\rho}_\Phi{\left( 1 \right)} },
\end{equation}
where $H_0$ is present Hubble rate, $\Omega_{\Lambda0}$ is present dark energy density, $\Omega_{\gamma0}$ is presented photon density, $\Omega_{\nu0}$ is present neutrino density, $\Omega_{b0}$ is present baryon density, $\Omega_{\text{CDM}0}$ is present cold dark matter density, $\Omega_{k0}$ is present curvature, and $\Omega_{\Phi0}$ is present density of species $\Phi$. As we need to ensure that the observed relic density is satisfied, we set $\Omega_{\text{CDM}0} \approx 0.25 - \Omega_{\Phi0}$; thus, we essentially have a two-component dark sector, with some heavy cold species and non-cold species $\Phi$. 

The corresponding evolution equations for the baryon, CDM, photon, and neutrino sectors are well known and can be found in Ref.~\cite{Callin2006}. However, in our case, we also have an extra collisionless evolution equation for the species $\Phi$, which, upon the Legendre series expansion, gives us a system of evolution equations for the expansion coefficients $\omega_i$ 
\begin{align}
\frac{d \omega_0}{dx} &= - \frac{q c k}{\mathcal{H} \epsilon} \omega_1 + \frac{d \phi}{d x} \frac{d \text{ln}{f_\Phi}}{d \text{ln}{q}}, \label{eqn:PhiTower0} \\
\frac{d \omega_1}{dx} &= \frac{q c k}{3 \mathcal{H} \epsilon} \left( \omega_0 - 2 \omega_2 \right) - \frac{\epsilon c k}{3 q \mathcal{H}} \psi \frac{d \text{ln}{f_\Phi}}{d \text{ln}{q}}, \label{eqn:PhiTower1} \\
\frac{d\omega_l}{dx} &= \frac{q c k}{\left( 2 l + 1 \right) \mathcal{H} \epsilon} \left( l \omega_{l -1 } - \left( l -1 \right) \omega_{l+1} \right) \text{ } \forall \text{ } l \in \left[ 2, l_\text{max} \right), \label{eqn:PhiTowerl} \\
\frac{d\omega_{l_\text{max}} }{dx} &= \frac{q c k}{\mathcal{H} \epsilon} \left( \omega_{l_\text{max} - 1}  - \frac{l_\text{max} + 1}{k \eta} \omega_{l_\text{max}} \right), \label{eqn:PhiTowerlmax}
\end{align}
where we have introduced a new variable $x = \text{ln}{\left( a \right)}$, $l_\text{max}$ is the value of $l$ at which the Legendre series is truncated, $c$ is the speed of light, $\mathcal{H}{\left( x \right)} = a{\left(x \right)} H{\left( a{\left( x \right)} \right)}$, $\epsilon{\left( x \right)} = \sqrt{q^2 + m_\Phi^2 a{\left( x \right)}^2}$, and $\eta{\left( x \right)}$ is the conformal time defined in terms of variable $x$ as 
\begin{equation}
    \eta{\left( x \right)} = \int_{-\infty}^x \frac{c dy}{\mathcal{H}{\left( y \right)}},
\end{equation}
and $k$ is the scale of interest. Each expansion coefficient $\omega_l$ satisfies the initial conditions 
\begin{align}
    \omega_0 &= \frac{1}{2} \psi \frac{d \text{ln}{f_\Phi}}{d\text{ln}{q}}, \\
    \omega_1 &= - \frac{\epsilon k \eta}{6q} \psi \frac{d \text{ln}{f_\Phi}}{d \text{ln}{q}}, \\
    \omega_2 &= - \frac{k^2 \eta^2}{30}   \psi \frac{d \text{ln}{f_\Phi}}{d \text{ln}{q}}, \\
    \omega_l &= \frac{q c k}{\left( 2 l + 1 \right) \epsilon \mathcal{H}} \omega_{l-1}, \text{ } \forall \text{ } l \in \left[3, l_{\text{max}} \right].
\end{align}
For the distribution function $f_\Phi$ sampled in the discrete bins of today's momentum $q$, we have the set of evolution equations \cref{eqn:PhiTower0,eqn:PhiTower1,eqn:PhiTowerl,eqn:PhiTowerlmax} for each value of $q$. In addition, the perturbations of metric $\phi$ and $\psi$ evolve according to 
\begin{align}
    \frac{d \phi}{d x} &= \psi - \frac{c^2 k^2}{3 \mathcal{H}^2} \phi + \frac{H_0^2}{2 \mathcal{H}^2} \left[ \frac{\Omega_{\text{CDM}0} \delta_{\text{CDM}}}{a} + \frac{\Omega_{b0} \delta_b}{a} + \frac{4 \Omega_{\gamma0} \Theta_0}{a^2} + \frac{4 \Omega_{\nu0} \mathcal{N}_0}{a^2} + \frac{\Omega_{\Phi0} a^2 \delta_\Phi \bar{\rho}_\Phi{\left( a\right)}}{ \bar{\rho}_\Phi{\left( 1 \right)}} \right], \label{eqn:PotentailEqn1} \\
    \psi &= - \phi - \frac{12 H_0^2}{c^2 k^2 a^2} \left[ \Omega_{\gamma0} \Theta_2 + \Omega_{\nu}{\mathcal{N}_2} + \frac{1}{2} \Omega_{\Phi0} a^4 \sigma_\Phi \frac{\bar{\rho}_\Phi{\left( a\right)}}{ \bar{\rho}_\Phi{\left( 1 \right)}} \right], \label{eqn:PotentialEqn2}
\end{align}
where $\Theta_l$ and $\mathcal{N}_l$ are the expansion coefficients in the Legendre series of the photon and neutrino sectors, respectively, $\delta_{\text{CDM}}$ and $v_{\text{CDM}}$ are the density and velocity perturbations of the CDM sectors, and $\delta_b$ and $v_b$ are the density and velocity perturbations of the baryon sector. Metric perturbations satisfy the following initial conditions
\begin{align}
    \psi &= \frac{1}{\frac{3}{2} + \frac{2}{5} f_\nu}, \\
    \phi &= - \left( 1 + \frac{2}{5} f_\nu \right) \psi,
\end{align}
where $f_\nu = \Omega_{\nu0} / \left( \Omega_{\gamma0} + \Omega_{\nu0} \right)$. To obtain the expression for the matter power spectrum, we need to take the Newtonian limit of the Einstein equation. A good place to start is by rewriting \cref{eqn:PotentailEqn1} in terms of the cosmic time $t$ and the scale factor $a{\left( t \right)}$ as 
\begin{equation}
    \frac{a}{\dot{a}} \dot{\phi} = \psi - \frac{c^2 k^2}{3 \dot{a}^2} \phi + \frac{H_0^2}{2 \dot{a}^2 a} \Omega_{M0} \Delta_M, \label{eqn:EquationNewtonLimit1}
\end{equation}
where $\dot{f}$ denotes the derivative of function $f$ with respect to the cosmic time $t$. As we are interested at present day matter power spectrum, we have also taken a matter-dominated limit by setting $\Omega_{\gamma0} = 0$, $\Omega_{\nu0} = 0$, and introducing a relative matter perturbation $\Delta_M$ defined as $\Omega_{M0} \Delta_M = \Omega_{\text{CDM0}} \delta_{\text{CDM}} + \Omega_{b0} \delta_b + \Omega_{\Phi0} \delta_\Phi a^3 \bar{\rho}_\Phi{\left( a \right)} / \bar{\rho}_\Phi{\left( 1 \right)}$ with $\Omega_m = \Omega_{\text{CDM0}} + \Omega_{b0} + \Omega_{\Phi0}$. Noting that \cref{eqn:PotentialEqn2} has no dependency on $\dot{a}$, we can trivially take a Newtonian $\dot{a} \rightarrow 0$ limit of \cref{eqn:EquationNewtonLimit1} resulting in 
\begin{equation}
    \frac{H_0^2}{2 a} \Omega_{M0} \Delta_M - \frac{c^2 k^2}{3} \phi = 0,
\end{equation}
which can be rearranged to yield a solution for the matter perturbations
\begin{equation}
    \Delta_M{\left( a, k\right)} = \frac{2a}{3} \left( \frac{c k}{H_0} \right)^2 \frac{\phi{\left(a, k \right)}}{\Omega_{M0}},
\end{equation}
which can be directly related to the present-day matter power spectrum at scale $k$ through the definition of the matter power spectrum
\begin{equation}
    P{\left(a, k \right)} =  \abs{\Delta_{M}{\left(a , k \right)}}^2 \mathcal{P}{\left( k \right)}, \label{eqn:PowerSpectrumDef}
\end{equation}
where $\mathcal{P}{\left( k \right)}$ is the proportionality factor dependent upon the inflation model used.

\bibliographystyle{apsrev4-1.bst}

\bibliography{main}

\end{document}